\documentclass[11pt,english]{article}
\usepackage[T1]{fontenc}
\usepackage[latin9]{inputenc}
\usepackage{amsmath}
\usepackage{graphicx}
\usepackage{amssymb,booktabs}
\usepackage[arrow, matrix, curve]{xy}

\usepackage[footnotesize]{caption} 

\makeatletter

\usepackage{epsfig}\usepackage{graphics}

\def\0{{\bf 0}}

\def\phi{\varphi}

\def\T{\T}

\catcode`@=11 \@addtoreset{equation}{section} \catcode`@=12

\usepackage{babel}

\begin{document}


\title{Hidden scaling patterns and universality\\  in written communication 
}
 
\author{M. Formentin
\thanks{Dipartimento di Fisica ``Galileo Galilei'', Universit\`a degli studi di Padova, Via Marzolo 8, 35131 Padova, Italy, \texttt{marco.formentin@rub.de
} %
},\,
A. Lovison 
\thanks{Dipartimento di Matematica, Universit\`a di Padova, via Trieste 63, I-35121 Padova, Italy, \texttt{lovison@math.unipd.it} 
},\, 
A. Maritan
\thanks{Dipartimento di Fisica ``Galileo Galilei'', Universit\`a degli studi di Padova, Via Marzolo 8, 35131 Padova, Italy, \texttt{maritan@pd.infn.it} %
}\,
and
G. Zanzotto
\thanks{Dipartimento di Psicologia Generale,  Universit\`a di Padova, via Venezia 12, I-35131 Padova, Italy, \texttt{zanzotto@dmsa.unipd.it
} %
}}

\maketitle

\begin{abstract} 
The temporal statistics exhibited by written correspondence appear to be media dependent, with features which have so far proven difficult to characterize. We explain the origin of these difficulties by disentangling the role of spontaneous activity from decision-based prioritizing processes in human dynamics, clocking all waiting times through each agent's `proper time' measured by activity. This unveils the same fundamental patterns in written communication across all media (letters, email, sms), with response times displaying truncated power-law behavior and average exponents near $-\frac{3}{2}$. When standard time is used, the response time probabilities are theoretically predicted to exhibit a bi-modal character, which is empirically borne out by our new years-long data on email. These novel perspectives on the temporal dynamics of human correspondence should aid in the analysis of interaction phenomena in general, including resource management, optimal pricing and routing, information sharing, emergency handling.
\end{abstract}



\section{Introduction} \label{sect:intro}

Remarkable statistical regularities observed in human and animal dynamics have attracted much attention in recent years \cite{nakamura2008,HANAI:kx,Henderson:2001yq,Wang:2010vn,wang2012random,Dezso:2006kx,Goncalves:2008yq,Gao:2010fj,topiAmos,Vazquez:2006zr,edwards2007,Rybski2009,antenodo2009,premananda2011}. A particularly interesting and studied case is given by written communication, which, whether on paper (`letters'), or in electronic form (`email'), is a most fundamental human activity, sustaining and giving the tempo to much of our civilization's advance \cite{Barabasi:2005yq,Vazquez:2006zr,Kossinets:2006vn,Malmgren:2008ys,Eckmann:2004vn,Malmgren:2009rt,Oliveira:2005fk,zhao2011,quwang2011,stouffer2005comments}. In recent times short-text messaging (`sms') has also been added to the repertoire of media through which humans intensely communicate with each other in writing \cite{WuZhou2010}.

\begin{figure}[p]
\centerline{\scalebox{0.75}{
\begin{picture}(420,600)(0,0)
\put(0,175){\includegraphics[width=150mm]{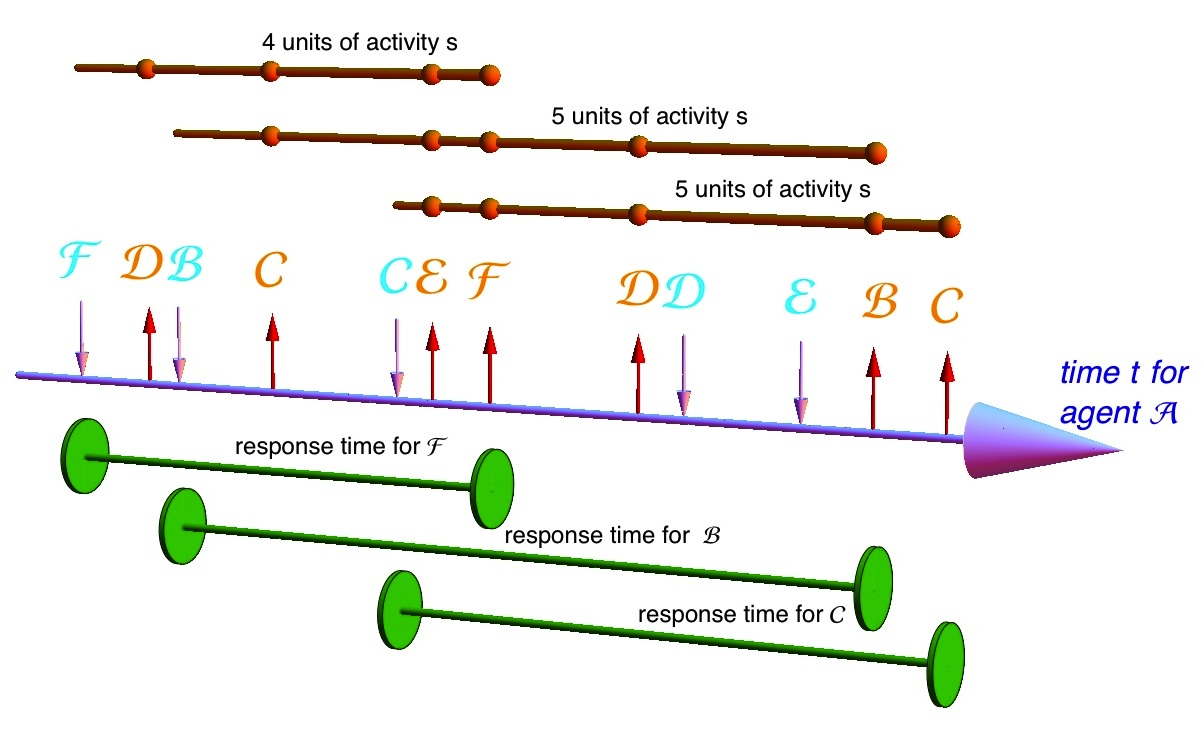}}
   \put(150,450){\includegraphics[width=70mm]{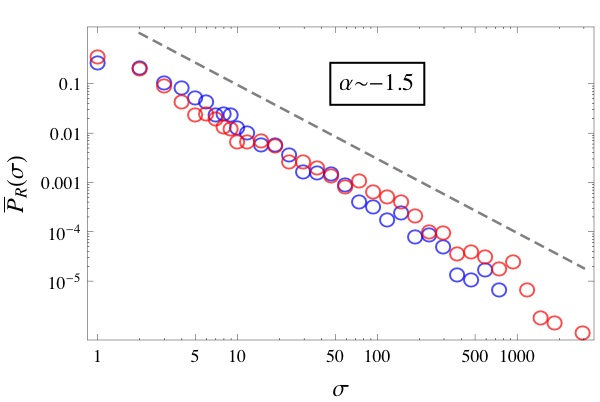}}
  \put(20,0){\includegraphics[width=70mm]{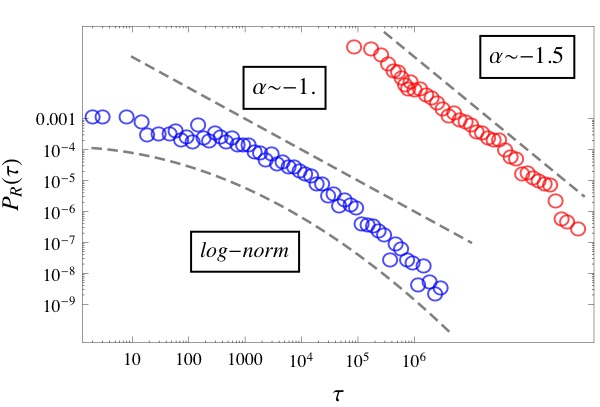}}
  \thicklines
  \put(210,455){\vector(-1,-3){15}}
  \put(150,135){\vector(1,4){16}}
  \put(0,115){$(a)$}
  \put(125,562){$(b)$}
\end{picture}
}}
\caption{Two clocks for written correspondence. Representation of the communication activity along the axis of time $t$ for an agent $\mathcal{A}$. Arrows pointing into the $t$ axis mark incoming messages from the indicated agents $\mathcal{B}$, $\mathcal{C}$, etc., arrows pointing out of the $t$ axis mark response messages to the same agents. The intervals between such arrows define the inter-event times (IETs) of agent $\mathcal{A}$. The response times (RTs) of $\mathcal{A}$ are defined as shown, either clocked through time $t$ (all measured in seconds), or through the activity parameter $s$ which counts the number of outgoing messages from $\mathcal{A}$ (see also the Supporting Information). The associated RT probability distributions are denoted by $P_{R}(\tau)$ and  $\bar{P}_{R}(\sigma)$ when clocked respectively through $s$ or $t$ (with $\tau = \Delta t$ and  $\sigma = \Delta s$). The RT distributions in terms of $t$ are non-universal, as they depend on the communication medium and the agent, see the lower diagram $(a)$, showing the $t$-clocked RTs of representative agents communicating through letters (red) and email (blue). In contrast, we find that the same RTs, when clocked through $s$, give distributions as in the upper diagram $(b)$, which are almost superposable power laws following eq.~[\ref{PL_troncate}], with individual exponents $\alpha$ on average near $-\frac{3}{2}$ for all media (letters, email, sms).
}
\label{cartoon}
\end{figure}
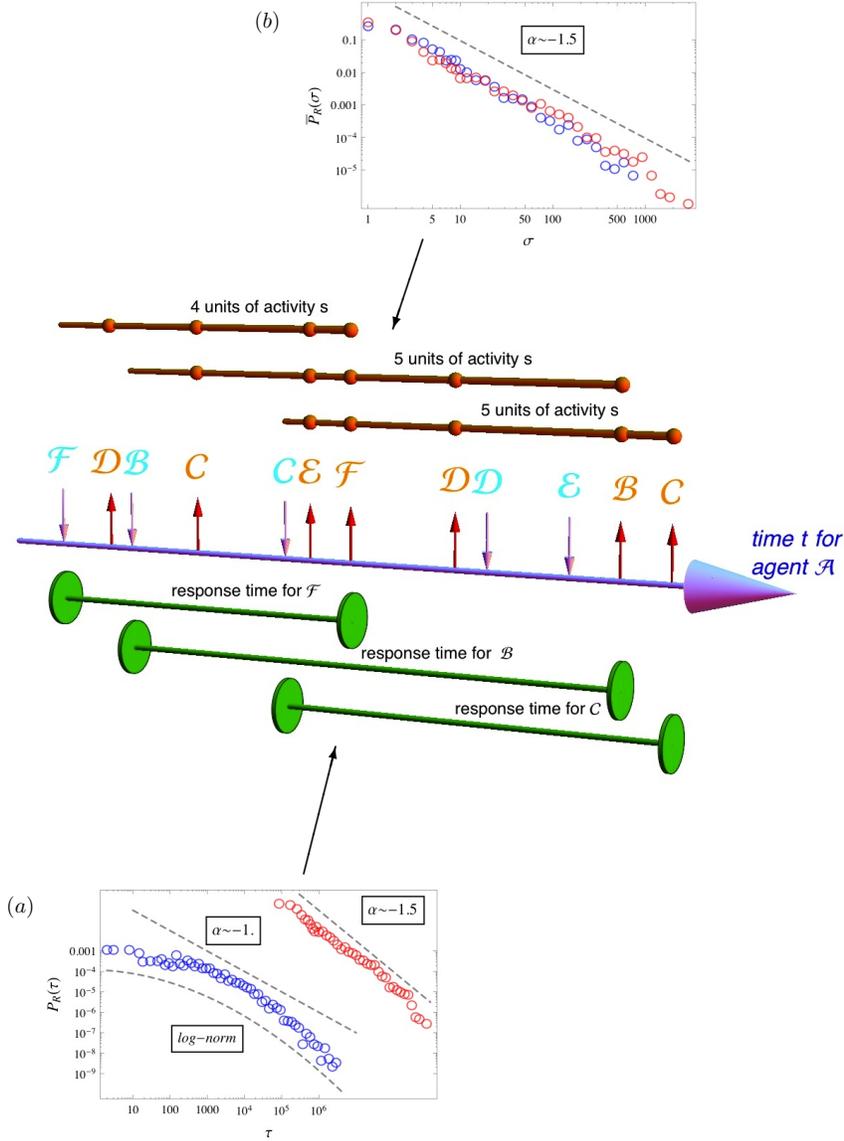

A main feature of interactive processes such as written correspondence is that, regardless of medium, the behavior and temporal dynamics of any agent $\mathcal{A}$ are characterized by two distinct waiting times, i.e.~response times (RTs) and inter-event times (IETs), schematically represented in Fig.~\ref{cartoon}; see also the Supporting Information (SI) for precise definitions.  We denote the probability distributions of RTs and IETs respectively by $P_{R}(\tau)$ and $P_{I}(\tau)$, where $\tau = \Delta t \in \mathbb{N}^+$ is the length of time intervals (with time $t$ measured in days for letters, and seconds for email and sms).
A better understanding of the mechanisms at the basis of written communication thus entails the analysis of these waiting times within large-scale interaction networks whose overall dynamics is largely unknown. During the last decade these and related questions have attracted the attention of a research community going from mathematics, to physics to sociology, whose studies, grounded on a number of databases which collect basic empirical information on communication events, have begun to clarify some basic facts on the behavior of such networks and the agents in it. In the SI we give details about the communication datasets used for the present work (denoted DL1, DE1, etc., see Table~\ref{table_exponents}), which include data previously available on written correspondence (letters, email, sms), as well as two new long-term email datasets collected for the present study.

\section{State of the art on time distributions and controversy} 

The first notable observation derived from the analysis of the empirical data is that events for all communication media occur in a highly intermittent fashion, with time fluctuations producing heavy-tailed distributions for both $P_{I}(\tau)$ and $P_{R}(\tau)$. The characterization of these statistics has been strongly debated, as they appear to depend on the medium (letters, email, sms) and lack universal features \cite{Barabasi:2005yq,Vazquez:2006zr,Malmgren:2008ys,Eckmann:2004vn,Malmgren:2009rt,WuZhou2010,JohansenCommento2006,Oliveira:2005fk,malmgremCritica2006, quwang2011, stouffer2005comments}, although the investigation in Ref.~\cite{Malmgren:2009rt} led to a form of universality for the IETs in letters and emails.  In spite of earlier indications of scaling for the empirical distributions $P_{R}(\tau)$ with two different exponents, $-1$ and $-\frac{3}{2}$, respectively in email and letters \cite{Barabasi:2005yq,Vazquez:2006zr,Eckmann:2004vn,JohansenCommento2006,Oliveira:2005fk,quwang2011},
 the scaling nature and general features of $P_{R}(\tau)$ for email are still contrastingly judged \cite{malmgremCritica2006}. Different priority queueing models have also been used to account for these controversial observations, producing power-law behavior for $P_{R}(\tau)$ with theoretical exponents $-1$ or $-\frac{3}{2}$  (see Refs.~\cite{Barabasi:2005yq,Vazquez:2006zr,cobham, Abate1996,Grinstein2008}), as well as exponents varying in a range from $-1$ to under $-2$ (Refs.~\cite{Barabasi:2005yq,IAT1,IAT2,blanchard2007,min2009,cho2010,kim2010,crane2010,mailart2011,saichev2010,jo2011}).
 



\section{Re-clocking the probability distributions through activity} 

To shed light on these poorly understood aspects of written communication, we disentangle from the overall time dynamics of a given agent $\mathcal{A}$ the contributions due to $\mathcal{A}$'s spontaneous inter-event pauses. To do this we introduce the parameter $s \in \mathbb{N}^+$ which counts the number of $\mathcal{A}$'s outgoing communication events (a measures of $ \mathcal{A}$'s activity), so that each increase by one unit for $s$ corresponds to an IET for $\mathcal{A}$, see Fig.~\ref{cartoon}. The probability densities for both the RTs and IETs, which characterize $\mathcal{A}$'s behavior, can be computed in terms of $\sigma = \Delta s$ in place of $\tau = \Delta t$. In analogy to similar clocking alternatives arising for instance in special relativity, the parameter $s$ can be interpreted, up to a suitable scale factor, as an agent's `proper time'; the introduction of $s$ bears also a relation to the `events per active interval' considered for different purposes in~\cite{Malmgren:2009rt}. We denote respectively by $\bar{P}_{R}(\sigma)$ and $\bar{P}_{I}(\sigma)$ the $s$-clocked probability distributions, and notice that the $s$-clocked IET distribution $\bar{P}_{I}(\sigma)$ is trivially the same for all agents and media, being concentrated by definition at $\sigma=1$. See eq.~[\ref{relazione_distribuzioni}] below and the SI for details on the mathematical relation among the probabilities $P_{R}(\tau)$, $P_{I}(\tau)$, and $\bar{P}_{R}(\sigma)$.




\section{Power-law empirical probabilities after re-clocking} 

\begin{figure*}[p]
\hspace*{-20mm}
\begin{tabular}{ccc}
\mbox{letters}&\mbox{emails}&\mbox{sms}\\
 \includegraphics[width=55mm,clip=]{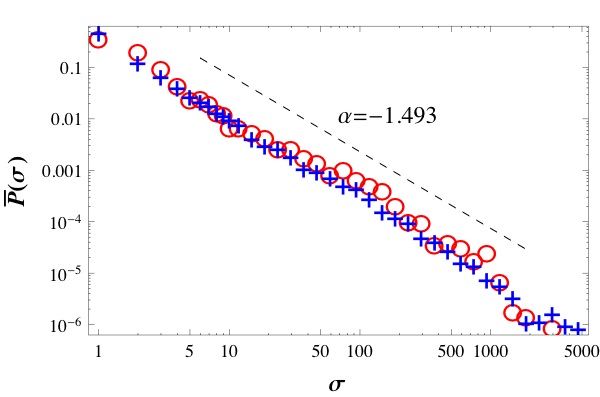} &
\includegraphics[width=55mm,clip=]{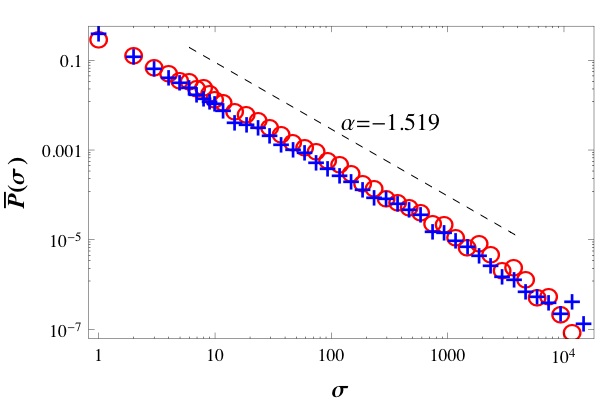} &
\includegraphics[width=55mm,clip=]{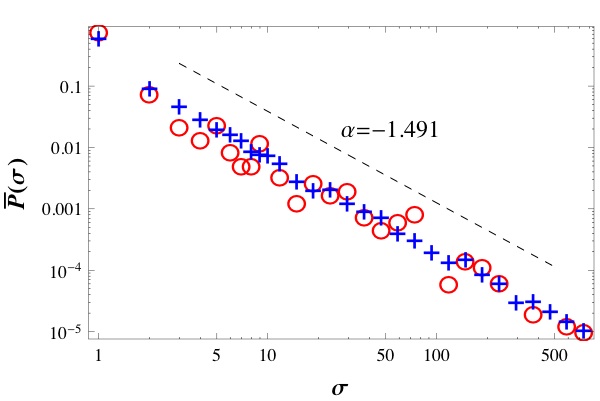}\\
 \includegraphics[width=55mm,clip=]{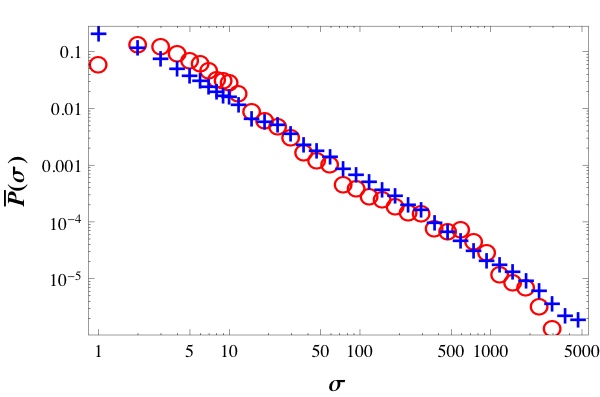}&
  \includegraphics[width=55mm,clip=]{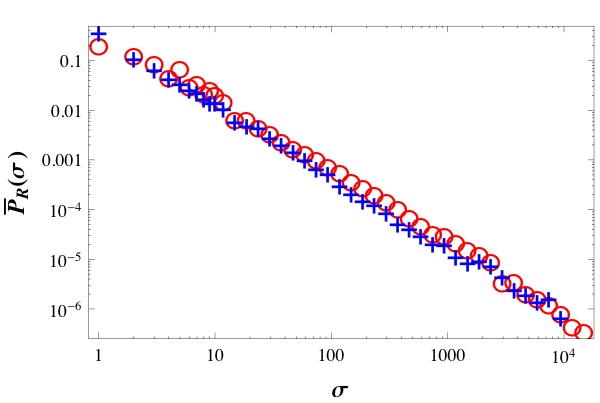}&
  \includegraphics[width=55mm,clip=]{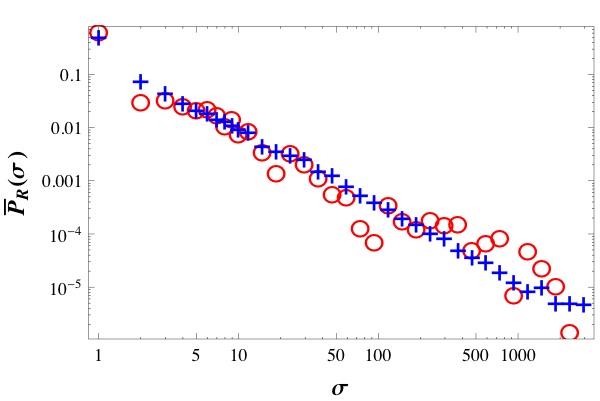} \\
 \includegraphics[width=55mm,clip=]{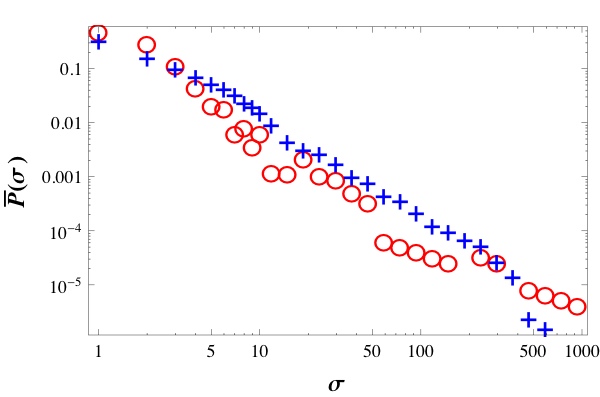} &
 $\includegraphics[width=55mm,clip=]{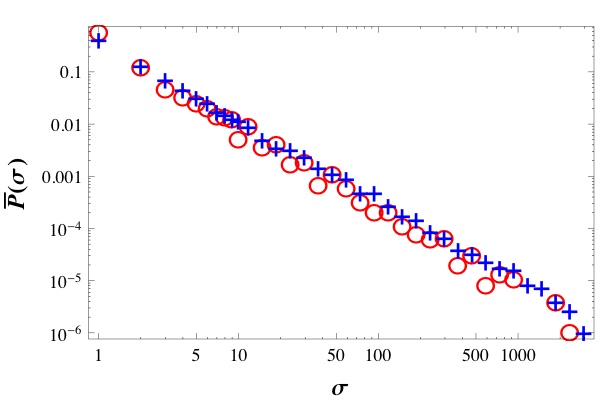}^{\scriptsize{(\star)}}$ &
 \includegraphics[width=55mm,clip=]{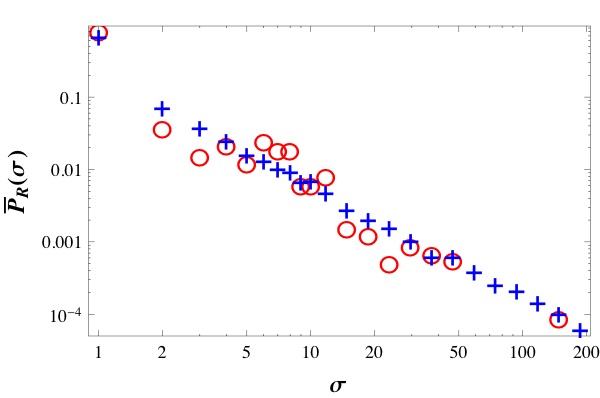}\\
(a) & (b) & (c)
\end{tabular}
\caption{ Log-log plots of the response-time probability densities $\bar{P}_{R}(\sigma)$ clocked through activity $s$, for three typical agents for each different written-communication medium (logarithmic binning \cite{stasapower}). Red circles indicate empirical data; blue crosses represent our model predictions.
(a) Letters: data from database DL1, on the correspondence of C. Darwin, A. Einstein, and S. Freud;
(b) email: data from typical agents in the long term databases DE1 and DE2 (the agent in DE2, with data spanning seven years, is marked by an asterisk);
(c) sms: data from typical agents in the database DS1 of Ref.~\cite{WuZhou2010}.
The probability densities for all media are very well fitted by the truncated power laws in eq.~[\ref{PL_troncate}] with individual exponents $\alpha$ as follows (going from top to bottom in each column): 1.493, 1.565, 1.886 (letters); 1.519, 1.604, 1.539 (email); 1.491, 1.215, 1.097 (sms). See Table~\ref{table_exponents} for information on the exponents in the various databases, and the SI for more statistics. The straight dashed lines in the top diagrams are drawn to guide the eye, with the indicated exponents.
}
\label{RT-TOCK_3/2}
\end{figure*}

Remarkably, we find that in all databases, across all media, the RTs of active agents, when clocked through activity $s$, are described by discrete exponentially-truncated power laws of the form
\begin{equation}
	\bar{P}_{R}(\sigma) \sim  {\sigma}^{\alpha} e^{-\frac{\sigma}{\lambda}},
	\label{PL_troncate}
\end{equation}
where $\alpha$ is the scaling exponent, $\lambda$ the cutoff parameter \cite{newman2009}.  A number of empirical distributions $\bar{P}_{R}(\sigma)$ as in eq.~[\ref{PL_troncate}], representative of the $s$-clocked RTs for each written communication medium (letters, email, sms) are shown in Fig.~\ref{RT-TOCK_3/2} (see the SI for more statistics). The individual exponents in the empirical RT distributions in eq.~[\ref{PL_troncate}] have average values close to $-\frac{3}{2}$ for all the three media, as detailed in Table~\ref{table_exponents}. The truncated scaling in eq.~[\ref{PL_troncate}] of $\bar{P}_{R}(\sigma)$ with exponents averaging near $-\frac{3}{2}$ can be clearly appreciated also the most active sms agents, despite their having comparatively much scarcer statistics than in email or letters. For email, these results on the scaling of $\bar{P}_{R}(\sigma)$ and its exponents are validated in agents across all the three independently-collected databases. We sampled the long-term email data through three-, six-, twelve- and eighteen-month windows within the total two-year period of dataset DE1, and for all window lengths we found great consistency in the obtained distributions of individual exponents, both within and across the three email datasets (see also Supporting Figs.~9-10(b)). 

Summarizing, while the waiting time distributions may vary across agents and media when expressed in terms of standard time $t$, all waiting times have quite the same medium-independent form when computed through proper time $s$, with a definite convergence of the exponents to average values near $-\frac{3}{2}$ in all media. This goes together with the universality of the $s$-clocked IET distributions $\bar{P}_{I}(\sigma)$, which are all concentrated at $\sigma=1$ as mentioned earlier. The introduction of the activity clocking thus emphasizes an intrinsic universal component underlying all written communication, partly obfuscated by the interaction with the spontaneous IETs, which are media- and agent-dependent. We discuss such universality more in detail below.

\begin{table*}[ht]
\caption{
On the left are indicated the databases analyzed in this work for the three written-communication media (letters, email, sms); see the SI for details. On the right are reported the corresponding exponents $\alpha$ computed for the empirical RT probabilities $\bar{P}_{R}(\sigma)$  in eq.~[\ref{PL_troncate}], clocked through activity $s$. Individual values of $\alpha$ are given for databases DL1, DE2; the average $\bar{\alpha}$ and standard deviation $\sigma$ of the distributions of individual exponents are indicated for the databases DE1, DE3, DS1. 
}


   \begin{tabular}{@{} c|c|c @{}} 
        \textbf{medium}  & \textbf{database} & \textbf{exponents}  \\
        \midrule
        letters \rule{15mm}{0mm}
        & \parbox{50mm}{\;  {\bf DL1}:
        agents CD, AE, SF} 
        & \parbox{50mm}{\;\; $\alpha_{CD}= 1.493 \pm 0.020$ \\\;\; $\alpha_{AE}= 1.565\pm 0.013$ \\\;\; $\alpha_{SF}= 1.886\pm 0.028$ }\\
        \midrule
        & \parbox{50mm}{\; {\bf DE1}:
        new two-year database 
} 
        & \parbox{50mm}{\;\; $\overline\alpha= 1.543 $ \\ \;\; $\sigma= 0.306  $}\\
       \cmidrule(l){2-3}
        email \rule{15mm}{0mm}
        & \parbox{50mm}{ \; {\bf DE2}:
        new very long term database, agents AL, AP, FC} 
        & \parbox{50mm}{\;\; $\alpha_{AL}=1.539\pm0.024$ \\ \;\; $\alpha_{AP}= 1.557\pm 0.011$ \\ \;\; $\alpha_{FC}= 1.478\pm 0.008$ }\\
       \cmidrule(l){2-3}
        & \parbox{50mm}{\; {\bf DE3}:
        three-month database from Ref.~\cite{Eckmann:2004vn}
} 
        & \parbox{50mm}{$\overline  \alpha= 1.562 $ \\ \;\; $\sigma= 0.366$}\\
        \midrule
        sms \rule{15mm}{0mm}
        & \parbox{50mm}{\; {\bf DS1}:
       one-month database from Ref.~\cite{WuZhou2010}
} 
        & \parbox{50mm}{$\overline\alpha= 1.447 $ \\ \;\; $\sigma= 0.444$}\\
        %
      \bottomrule
   \end{tabular}
   %
   \label{table_exponents}
\end{table*}

\begin{figure*}[h!]
\hspace*{-20mm}
\begin{tabular}{ccc}
\mbox{letters}&\mbox{emails}&\mbox{sms}\\
 \includegraphics[width=55mm,clip=]{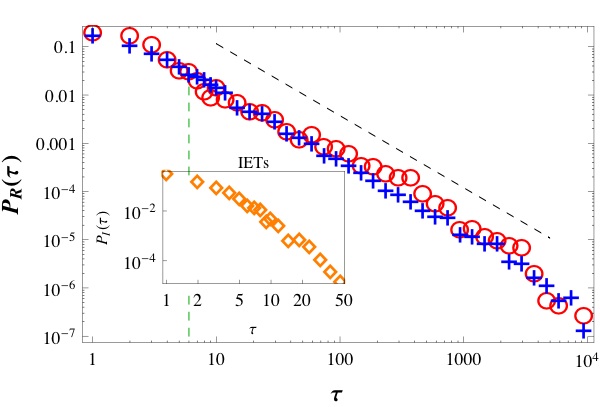}&
 \includegraphics[width=55mm,clip=]{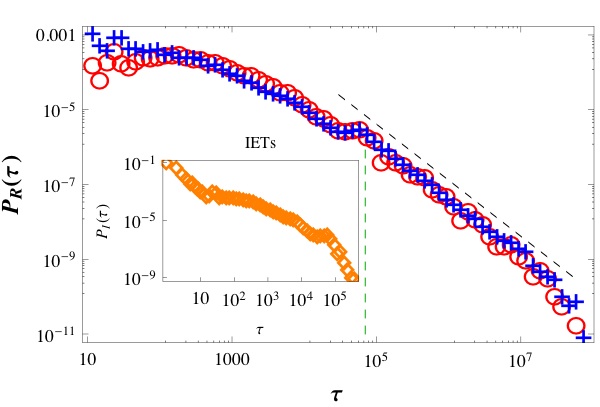}&
 \includegraphics[width=55mm,clip=]{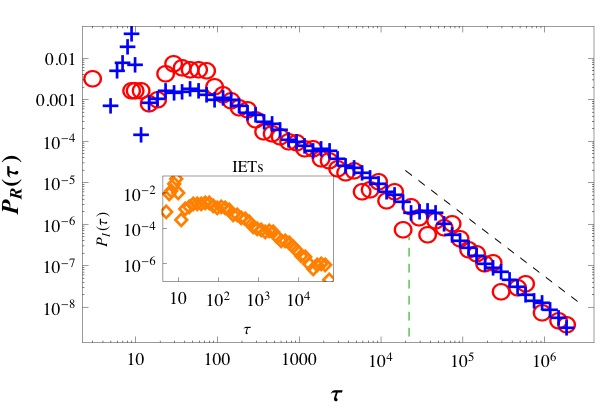}\\
 \includegraphics[width=55mm,clip=]{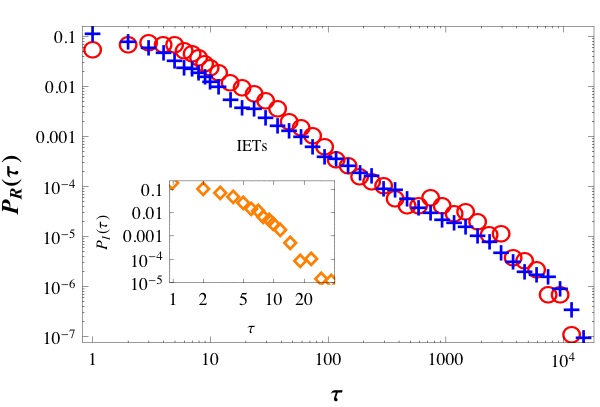}&
 \includegraphics[width=55mm,clip=]{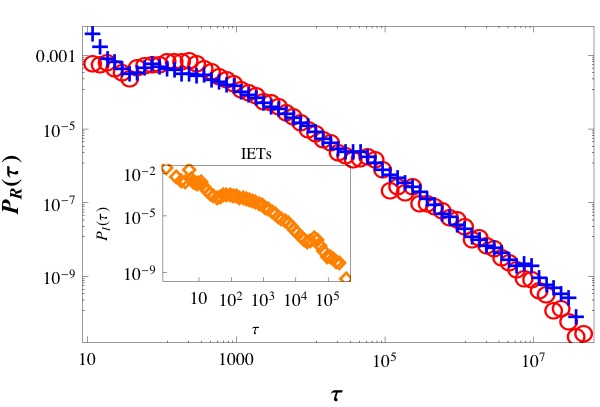}&
 \includegraphics[width=55mm,clip=]{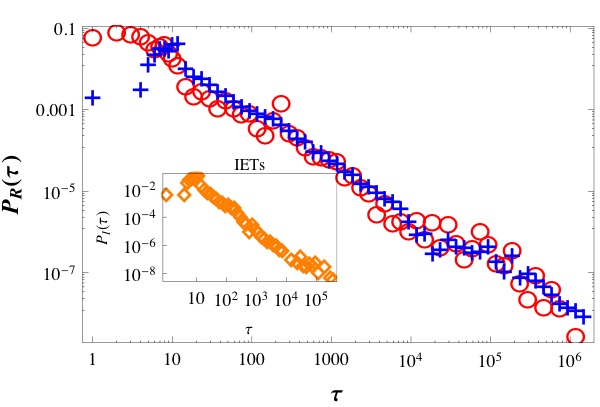} \\
 \includegraphics[width=55mm,clip=]{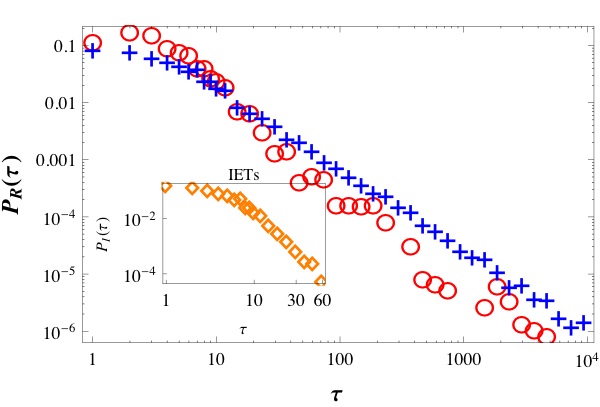} &
 $\includegraphics[width=55mm,clip=]{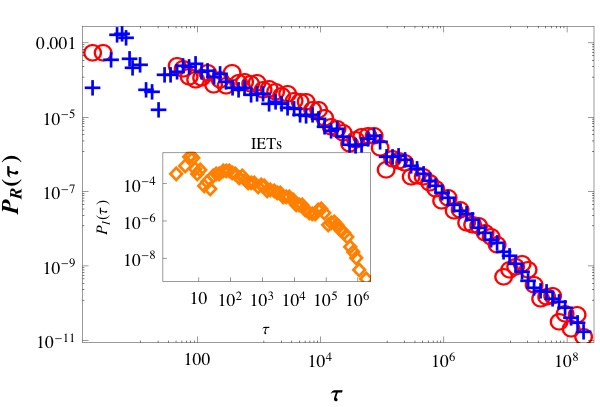}^{\scriptsize{(\star)}}$&
\includegraphics[width=55mm,clip=]{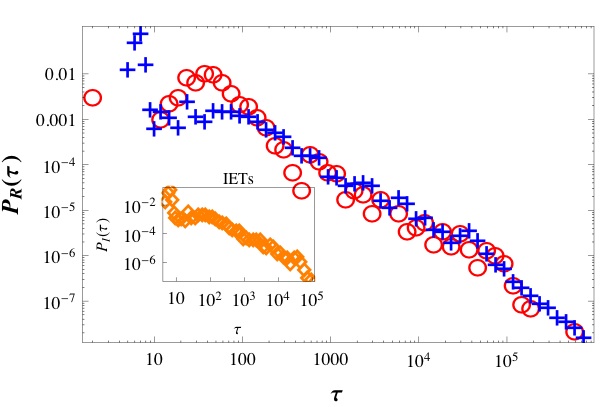}\\
(a) & (b) & (c)
\end{tabular}
\caption{Log-log plots of empirical response-time probability densities $P_{R}(\tau)$ clocked through standard time $t$ (in days for letters, and seconds for email and sms), relative to the same typical agents as in Fig.~\ref{RT-TOCK_3/2}, for all media (logarithmic binning~\cite{stasapower}). Red circles indicate empirical data; blue crosses represent computational predictions. See the SI for more statistics. As predicted (see the SI), $P_{R}(\tau)$ is affected, for small $\tau$, by the specific features of $P_{I}(\tau)$, while the tails of $P_{R}(\tau)$ for large $\tau$ follow power laws with the same exponents $\alpha$ as the associated $s$-clocked distributions $\bar{P}_{R}(\sigma)$, shown in Fig.~\ref{RT-TOCK_3/2}. The bi-modality in these $t$-clocked RT probabilities $P_{R}(\tau)$ is particularly evident in the RTs for email, in column (b).
Also following predictions (see the SI), the crossover in $P_{R}(\tau)$ occurs for $\tau \sim T_{I}$ (green dashed vertical lines), where $T_{I}$ is the characteristic time of the empirical IET distributions $P_{I}(\tau)$ of each agent, shown in the insets. Typical empirical values are $T_{I} \sim 10^4$-$10^5$ sec for email and sms, and $T_{I} \sim 5$-$10$ days for letters.
}
\label{tick_bimodal}
\end{figure*}




\section{Bi-modal empirical probabilities clocked through time} 

In the light of the above results on the $s$-clocked distributions $\bar{P}_{R}(\sigma)$, we can now better analyze the empirical $t$-clocked RT distributions $P_{R}(\tau)$ of written correspondence. For the same agents as in Fig.~\ref{RT-TOCK_3/2}, and for each medium (letters, email, sms), the individual $P_{R}(\tau)$ are represented in Fig.~\ref{tick_bimodal}, the insets showing the associated IET distributions $P_{I}(\tau)$. We see from Fig.~\ref{tick_bimodal} that the $t$-clocked RT distributions $P_{R}(\tau)$ do not scale, and exhibit complex, media-dependent characteristics (more statistics are reported in the SI). 

This behavior of $P_{R}(\tau)$ can be understood by considering that the $t$-clocked distribution $P_{R}(\tau)$ of any agent $\mathcal{A}$ can be retrieved in a natural way by compounding the IET probabilities $P_{I}(\tau)$, characterizing the spontaneous action of $\mathcal{A}$, back into the $s$-clocked RT power law $\bar{P}_{R}(\sigma)$ in eq.~[\ref{PL_troncate}], i.e.~by separating any two consecutive activities of $\mathcal{A}$ through random time intervals sampled from the IET distribution $P_{I}(\tau)$ of $\mathcal{A}$ (representative examples of IET distributions in the different media are shown in the insets of Fig.~\ref{tick_bimodal}). Specifically, let $N \sim \bar{P}_{R}(\sigma)$ and $\rho_{I}(h) \sim P_{I}(\tau)$, $h=1,2,\dots$, be independently-sampled random variables, with $N$ giving the number of activities between a message reception by $\mathcal{A}$ and the response to it; then, the $t$-clocked RTs for $\mathcal{A}$ are described by the compounding process
\begin{multline}
\label{relazione_distribuzioni}
\rho_R =\sum_{h=1}^N\rho_{I}(h)\;\; \mbox{ with law }\;\;
P_R(\tau) =\sum_{\sigma\geq 1}\mbox{Prob}\left(\sum_{h=1}^{\sigma}\rho_{I}(h)=\tau\right)
\bar{P}_{R}(\sigma).
\end{multline}
Numerical simulations confirm the above relation holds for the empirical distributions $P_{R}(\tau)$, $P_{I}(\tau)$, $\bar{P}_{R}(\sigma)$, indicating implicitly that correlations in the waiting times of human correspondence, if any, do not significantly affect the compounding of probabilities in eq.~[\ref{relazione_distribuzioni}]. This agrees with the results in \cite{anteneodo2010} indicating a lack of correlations within the IET statistics from the email data in \cite{Eckmann:2004vn}.

In the SI we show that the $t$-clocked RT distributions $P_{R}(\tau)$ in eq.~[\ref{relazione_distribuzioni}] result to have a {\it bi-modal} character when the IET distribution $P_I(\tau)$ is heavy tailed and $\bar{P}_{R}(\sigma)$ is scaling as in eq.~[\ref{PL_troncate}]. It results that, due to eq.~[\ref{relazione_distribuzioni}], for large $\tau$, $P_{R}(\tau)$ has power-law tails with the same exponent near $-\frac{3}{2}$ as $\bar{P}_{R}(\sigma)$, while, for small $\tau$, $P_{R}(\tau)$ is affected by the specific features of $P_{I}(\tau)$, see Supporting Fig.~8. The crossover in $P_{R}(\tau)$  occurs for $\tau$ of the order of the characteristic time $T_{I}  \sim \frac{<\tau^2>}{<\tau>}$ of the empirical IET distributions $P_{I}(\tau)$. In accordance to such prediction, we see in Fig.~\ref{tick_bimodal} that the empirical $t$-clocked RT probabilities about $P_{R}(\tau)$ do exhibit media-dependence with a complex, bi-modal behavior. The latter is particularly evident in the $P_{R}(\tau)$ distributions derived from the new long-term data on email, which span the largest number of decades in time, from seconds to several years (databases DE1 and DE2). The bi-modality of $P_{R}(\tau)$ likely led to the controversial conclusions earlier reported in the literature about the time statistics in email communication, which our analysis now contributes to clarify.
%
%
%
%

%

\section{Modeling and universal mechanism} 

To establish a theoretical basis for the above observations on the time patterns of written communication, we show that both the empirically reported $s$- and $t$-clocked statistics (Figs.~\ref{RT-TOCK_3/2}-\ref{tick_bimodal}) can be interpreted through priority queueing. We build on previous work about such modeling for human correspondence \cite{Barabasi:2005yq, cobham,Oliveira:2005fk,Vazquez:2006zr, Gabrielli2009, Gabrielli2009-1, Abate1996,Grinstein2008, IAT1, IAT2},
and demonstrate that we can obtain both the scaling distributions $\bar{P}_{R}(\sigma)$ in eq.~[\ref{PL_troncate}], as well as the bi-modal distributions $P_{R}(\tau)$ derived from eq.~[\ref{relazione_distribuzioni}], once the individual IETs and the message arrival times of each agent are suitably accounted for within a universal prioritization framework.

Let $\mathcal{A}$ be an agent with given IET empirical distribution $P_{I}(\tau)$ (see the examples in Fig.~\ref{tick_bimodal}), and assume for $\mathcal{A}$ an initial list of $L$ tasks, whose priorities $y$ are sampled from the uniform distribution on $[0,1]$ (consistent with the hypothesis that $\mathcal{A}$ is embedded in a complex communication network producing largely independent stimuli to $\mathcal{A}$). At each time step, corresponding to a unit increment of $\mathcal{A}$'s activity $s$, the highest priority task in the list is executed (a message replied), and $m$ new tasks are added to the list, each one with priority $y$ sampled as above. The number $m$ is derived at each time step by considering the empirical distribution of incoming messages to $\mathcal{A}$ between any two consecutive message activities of $\mathcal{A}$, the data typically giving $m > 1$. The numerical results for the $s$-clocked steady-state RT distribution $\bar{P}_{R}(\sigma)$ for this process are shown in Fig.~\ref{RT-TOCK_3/2}. For all media (letters, email, sms) the simulations follow extremely closely their empirically-observed counterparts, tracing power laws with the correct individual exponents even for values with large departures from the average near $-\frac{3}{2}$ (see the SI for details on the statistical analysis of the compatibility between numerical results and empirical data, according to \cite{newman2009, Rousseau:2000tk,Conover:1972vs,Arnold:2011ty}).

The model also accounts for the bi-modality of the $t$-clocked RT statistics of human correspondence, reported in Fig.~\ref{tick_bimodal}. The distribution $P_{R}(\tau)$ of each agent can again be derived from the computed $\bar{P}_{R}(\sigma)$ as in Fig.~\ref{RT-TOCK_3/2}, by separating, as in eq.~[\ref{relazione_distribuzioni}], the activity events of $\mathcal{A}$ through random time intervals sampled from the empirical IET distribution $P_{I}(\tau)$ pertaining to $\mathcal{A}$. The log-log plots of the distributions $P_{R}(\tau)$ so obtained are shown in Fig.~\ref{tick_bimodal}. We see that the numerical predictions are virtually indistinguishable from the empirical results for all media. 
This confirms that for active agents our approach consistently reproduces very well the empirical data for both the $s$- and $t$-clocked RT distributions across all media in a wide range of estimated exponents. See also Supporting Figs.~4-7.




\section{Conclusions} 

Our findings highlight the interplay between individual spontaneous activity (subsumed by the IET distributions) and universal decision-based processes (subsumed by task-prioritization) in the origin of the complex time patterns of written communication. We determine the role of both these factors in the generation of scaling $s$-clocked RT distributions $\bar{P}_{R}(\sigma)$ with exponents $\alpha$ distributed near $-\frac{3}{2}$, as well as  (through the compounding in eq.~[\ref{relazione_distribuzioni}]) in producing bi-modal $t$-clocked RT distributions $P_{R}(\tau)$, in very close accordance to empirical data in all media. This gives a novel perspective on the nature and features of the temporal inhomogeneities in human dynamics and their underlying mechanisms; in particular, our results refute earlier views on the media-dependent power-law or log-norm character of the $t$-clocked response functions for letters and email, bringing these two media within the same setting, with text messaging as well. 

Interestingly, we see that the power-law behavior in eq.~[\ref{PL_troncate}] does not arise when written communication occurs mostly in pairs, as analyzed in Ref.~\cite{WuZhou2010}, because in this case the $t$-clocked RTs and IETs are strongly correlated, i.e., the $s$-clocked RTs almost coincide with the $s$-clocked IETs, being both concentrated near $\sigma=1$. In contrast, human dynamics with large fluctuations and scaling statistics arises from the operation of complex interaction networks with rich-enough topologies. Prioritization processes then give average values near $-\frac{3}{2}$ to the emerging exponents $\alpha$, although the latter bear the signature of each agent's input from the network, as the individual deviations of $\alpha$ from $-\frac{3}{2}$ are shown by the model to be affected by the specific arrival-time statistics. To a lesser degree, the exponents may further be influenced by other factors, such as social structure, interest, habit, as discussed in Refs.~\cite{Kentsis:2006uq, blanchard2007,min2009,cho2010,kim2010,crane2010,mailart2011,saichev2010,jo2011}. 
While in our approach the IET distributions $P_{I}(\tau)$ of agents are derived from the empirical data, various avenues for a theoretical understanding of IETs can be considered, along the lines of Refs.~\cite{Vazquez:2005fk,Vazquez:2006zr,topiAmos, Malmgren:2008ys, impact2007,impact2009,habit2010}. The explicit IET fit proposed in Ref.~\cite{Malmgren:2008ys} could also be used in eq.~[\ref{relazione_distribuzioni}] to obtain a fully numerical reproduction of the empirical data.
This complements our insight into the dynamics of written correspondence as representing the wider network of human interactions, driven by distributed co-operative effects as well as deliberate vs.~spontaneous individual processes. The proposed methods may help uncover and analyze hidden patterns also in other contexts for the interactive dynamics of human and non-human agents alike.


%


\medskip

{\em \textbf {Acknowledgments:}}  We thank Drs.~J.-P.~Eckmann, M. Gravino, R.~D.~Malmgrem, J.~Oliveira, A.~Pellizzon, and three individual long-term email users, for providing us part of the communication data analyzed in this study. AM acknowledges the Cariparo Foundation for financial support.

\newpage

\part*{Supporting Information}
 

\section{Databases}

The databases for written correspondence analyzed in this study concern the three communication media: letters, email, text messages (sms). The collected data are in the form {\{\sc sender, receiver, timestamp\}}, where senders and receivers are conventionally numbered, and the timestamps are given in days for letters, and in seconds for emails and sms.


\subsection{Paper correspondence (letters)} \label{letters}

We have considered for letters the following {\bf Database DL1}, comprising the available life-time correspondence data for three well-known writers (see also \cite{Oliveira:2005fk, Vazquez:2006zr}):

\smallskip

\begin{itemize}

\item  C. Darwin, see \texttt{http://www.darwinproject.ac.uk/}

\smallskip

\item  A. Einstein, see \texttt{http://www.alberteinstein.info/}

\smallskip

\item  S. Freud, see \texttt{http://www.freud.org.uk/}

\end{itemize}


\subsection{Email}\label{longemaildata}

We have considered the following three databases for email:

\smallskip

\begin{itemize}

\item {\bf Database DE1}.~~This is a newly collected email database concerning the long-term activity of all the accounts belonging to, and interacting with, a Department of a large EU university, extending over a period of about two years. This dataset is available as a separate file in this Supporting Information.

\smallskip

\item {\bf Database DE2}.~~This is a newly collected email database comprising the very long-term email activity of three agents, extending over periods of five to nine years. This dataset is available as a separate file in this Supporting Information.

\smallskip

\item {\bf Database DE3}.~~This is the short term email database from \cite{Eckmann:2004vn}, comprising data referring to a EU university, covering a period of about three months.

\end{itemize}


\subsection{Text messages (sms)}\label{sms}

We have considered for sms the {\bf Database DS1} available from \cite{WuZhou2010}, comprising data on the accounts belonging to three Chinese companies, extending over a one-month period, see \texttt{adsabs.harvard.edu/abs/2010PNAS}




\section{Definitions and probability densities of IETs and RTs \label{definitions}}

\subsection{Inter-event times (IETs)} Referring to a communicating agent $\mathcal{A}$ in any of the above databases, the IETs are the time intervals $\tau := \Delta t$ (in seconds for email and sms, in days for letters) between two consecutive activity events of $\mathcal{A}$, i.e.~the time intervals separating the acts of sending two consecutive letters, emails, or sms by $\mathcal{A}$. The probability distribution of IETs is denoted by $P_{I}(\tau)$. By introducing the parameter $s \in \mathbb{N}^+$ counting the number of outgoing communication events (i.e.~the activity) of $\mathcal{A}$, we can compute the IETs of $\mathcal{A}$ also through $s$ (see Fig.~1 in the main text). For IETs by definition we have $\sigma := \Delta s \equiv 1$, so that the corresponding $s$-clocked IET probability distribution $\bar{P}_{I}(\sigma)$ is concentrated at 1, $\bar{P}_{I}(\sigma) \equiv \delta_1$, where $\delta$ is Dirac's delta distribution.

\smallskip

\subsection{Response times (RTs)}~~The RTs pertaining to agent $\mathcal{A}$ are the time intervals $\tau = \Delta t$  (in seconds for email and sms, in days for letters) separating the arrival of any message $\mathcal{M}$ from any agent $\mathcal{B}$ to $\mathcal{A}$, and the first ensuing message $\mathcal{M}$$'$ going from $\mathcal{A}$ to $\mathcal{B}$, independently of the subject or contents of $\mathcal{M}$ or $\mathcal{M}$$'$ (a response time for $\mathcal{A}$ may thus refer to the time taken for the actual reply of $\mathcal{A}$ to a message from $\mathcal{B}$, or also to the time taken by $\mathcal{A}$ to renew a perhaps interrupted correspondence interaction with $\mathcal{B}$). The RTs of $\mathcal{A}$ can also be defined through the activity parameter $s$ of $\mathcal{A}$ by counting the values $\sigma = \Delta s$ pertaining to the intervals between the same messages $\mathcal{M}$ and $\mathcal{M}'$ as above, i.e.~the number of outgoing messages from $\mathcal{A}$ intervening between $\mathcal{M}$ and $\mathcal{M}'$ (see Fig.~1 in the main text). The RT probability density of $\mathcal{A}$ can then be computed in terms of either $t$ or $s$, producing the distributions $P_{R}(\tau)$ and $\bar{P}_{R}(\sigma)$ respectively.
The relation among the distributions $P_{R}(\tau)$, $P_{I}(\tau)$, and $\bar{P}_{R}(\sigma)$ is discussed in Sect.~\ref{bimodal}.




\section{Data selection}

The databases above contain raw data on different communication technologies each exhibiting specific usage styles and problems, with marked differences across media, and presenting specific problems related to the length of the observation window in each database. We followed the basic criteria below to select relevant data or users in each database.




\subsection{Data selection in the letters database DL1}

Two main concerns regard this dataset: the first problem are missing letters, which are a virtual certainty for all the considered authors, especially in the first part of their lives. This can be checked by considering the very large intervals in the tail of the empirical IET distributions in the life-long data, which for all the three writers extend to the order of years. The second question is that of non-stationarity, due to increased average activity in roughly the second half of all the three writers' lives, as compared with their earlier years (see also \cite{Oliveira:2005fk, Malmgren:2009rt}). For these reasons we selected only the correspondence data concerning about the last thirty years in the life of each writer. This alleviates both the concerns regarding non-stationarity and missing data. We notice that S. Freud's correspondence has the lowest number of RTs, and the longest IETs (possibly indicating a higher number of missing letters) among all the three considered authors.




\subsection{Data selection in the email database DE1}

In this new long-term email database we have first considered the 500 agents with the largest number of outgoing messages, and having a ratio $r:=\frac{\#{\text{incoming}}}{\#{\text{outgoing}}}$ in a wide interval
as in the earlier analysis of dataset DE3 performed in \cite{Malmgren:2008ys}.
From these, we have extracted the 300 agents with the largest number of question-reply pairs (this gives a set of agents with at least 390 RTs each, a large percentage of which have in the order of a few thousand RTs).




\subsection{Data selection in the email database DE2}

All the data were used in this new very long-term email database, for all three agents.




\subsection{Data selection in the email database DE3}

In this short-term email dataset, taken from \cite{Eckmann:2004vn}, we have first considered the 400 agents with the largest number of outgoing messages, and ratio $r$ as above.
From this subset of agents we have extracted those with at least 100 question-reply pairs each, in order to obtain reliable statistics from the data. This results in a set of active agents typically having a few hundred RTs each (i.e.~about 10$\%$ of the RTs if compared to the agents in the new email database DE1).
%
%




\subsection{Data selection in the sms database DS1}

In this short-term sms dataset, taken from \cite{WuZhou2010}, we have first considered the 500 agents with the largest number of outgoing messages, and ratio $r$ as above.
Furthermore, we have selected the sms accounts for which no more than 30\% of total traffic is directed to the most active correspondent (this percentage is about $10\%$ for emails and letters).%
\footnote{~This further selection in the sms database DS1 is necessary in order to obtain a set of agents with a significant variety of correspondents, comparable to those of letters and email. This is because in DS1 there are a majority of active accounts which interact in isolated or almost isolated pairs, writing only, or almost only, to each other. Indeed (see also \cite{WuZhou2010}) a large fraction of the active users in DS1 have highly polarized communications, writing prominently to one correspondent, and to few others: about 10\% of users have a single correspondent; about 50\% of the accounts have 90\% of total traffic directed to a single correspondent; for about 80\% of the accounts the majority ($> 50\%$) of total traffic occurs with one given correspondent. Thus, overall, a high focus in destination is observed for sms users in database DS1, unlike with the two other communication media (typically, less than 20\% of total traffic from letter or email writers is directed to their most active correspondent). Since we are interested in the common features of all media for written correspondence, typically occurring in communication networks with high connectivity, we have extracted from dataset DS1 the active users whose traffic involves a sufficiently high number of correspondents. We notice the set of sms users so selected lies at the opposite end of the polarization spectrum as compared to the set of sms users considered in \cite{WuZhou2010}, whose analysis and modeling specifically focus on the behavior of highly polarized sms accounts. For these essentially pair-wise-interacting agents the $t$-clocked IET and RT statistics (which show a form of bi-modality, see \cite{WuZhou2010}) are strongly correlated, being almost identical to each other in the limit of a single correspondent. Accordingly, the RT intervals $\sigma =\Delta s$ of such agents are always small, with $\sigma = \Delta s \sim 1$ in the case of a single correspondent, their $s$-clocked RT distributions $\bar{P}_{R}(\sigma)$ being concentrated near $\sigma=1$, i.e., very similar to the associated $s$-clocked IET distributions $P_{I}(\sigma) \equiv \delta_1$.
\label{footnote_pairs}
}
Finally, from this set we have extracted the agents with at least 100 question-reply pairs, resulting in a set of sms agents typically having a few hundred RTs, comparable to the above dataset DE3. As with the latter, due to the short-time window also these DS1 agents have in general much scarcer statistics (about 10$\%$) if compared to agents in the long-term email databases DE1.
%
%





\section{RT statistics clocked through activity $s$}





\begin{figure}
\begin{center}
\textsc{Response-time distributions of email clocked through activity $s$}
\end{center}
\hspace*{-20mm}
\begin{tabular}{ccc}
 \includegraphics[width=50mm,clip=]{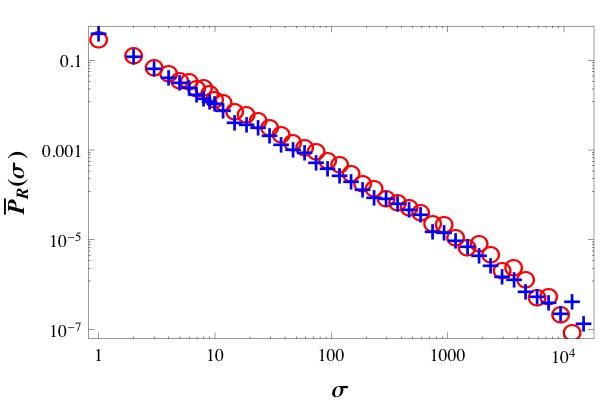} &
\includegraphics[width=50mm,clip=]{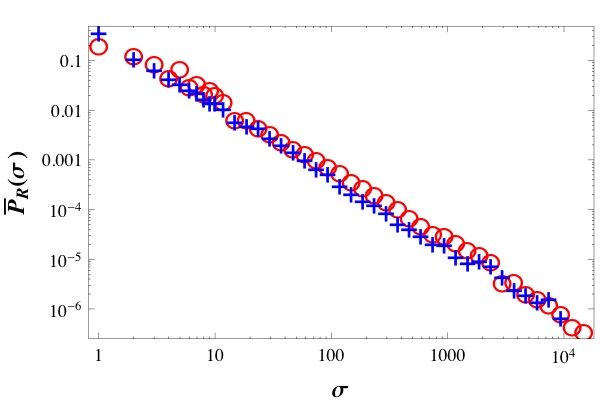} &
\includegraphics[width=50mm,clip=]{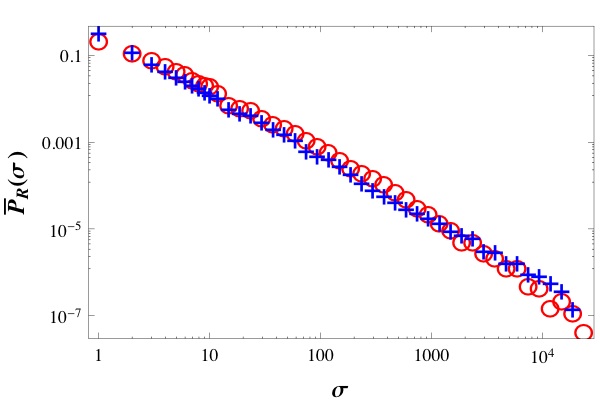}\\
 \includegraphics[width=50mm,clip=]{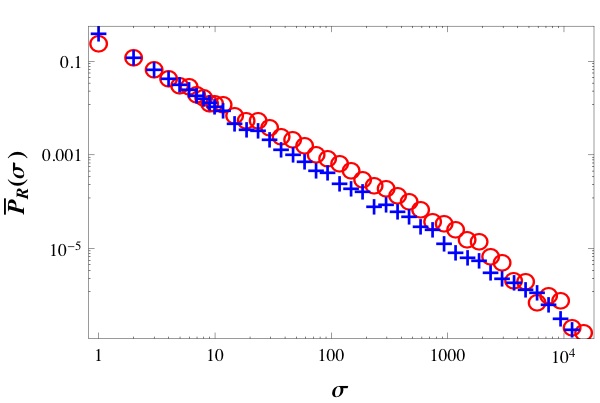} &
\includegraphics[width=50mm,clip=]{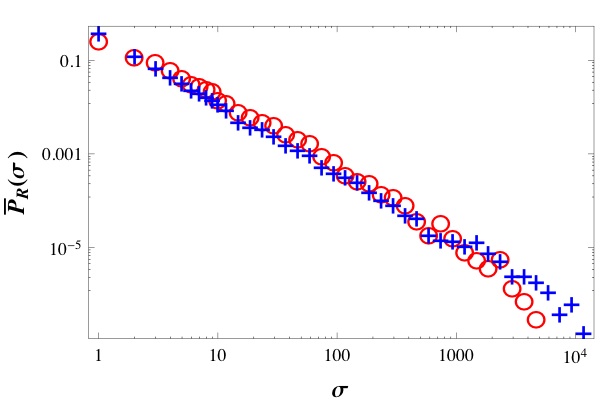} &
\includegraphics[width=50mm,clip=]{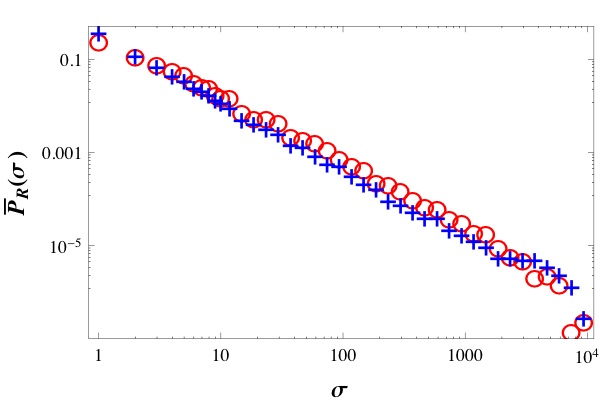}\\
 \includegraphics[width=50mm,clip=]{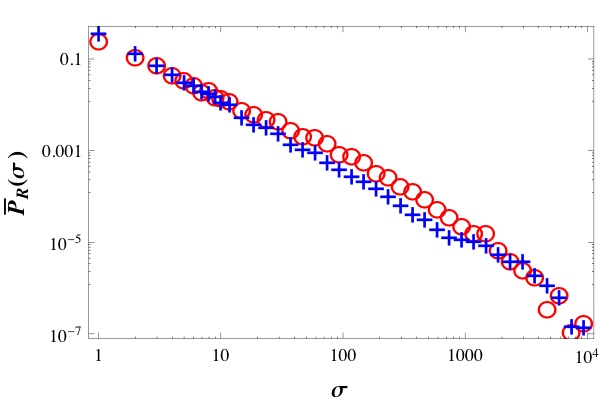} &
\includegraphics[width=50mm,clip=]{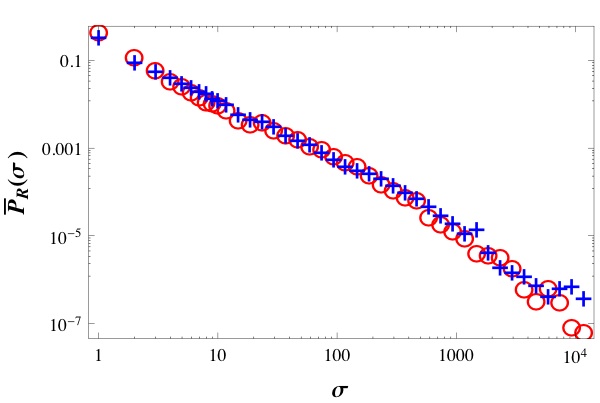}&
\includegraphics[width=50mm,clip=]{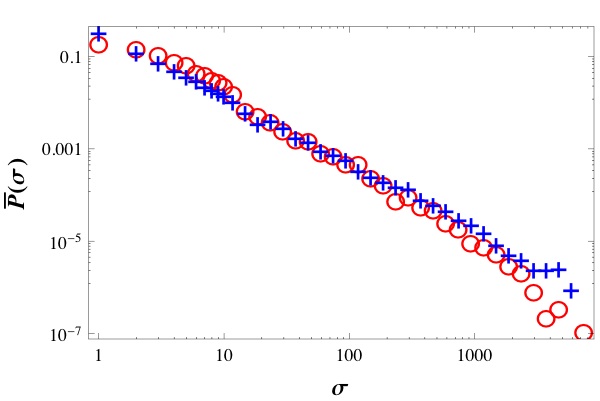} \\
\includegraphics[width=50mm,clip=]{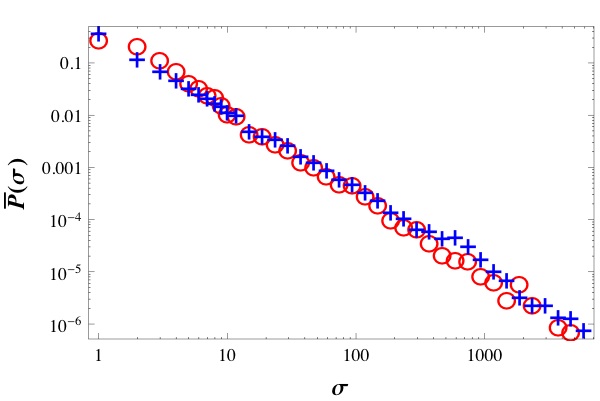}&
 \includegraphics[width=50mm,clip=]{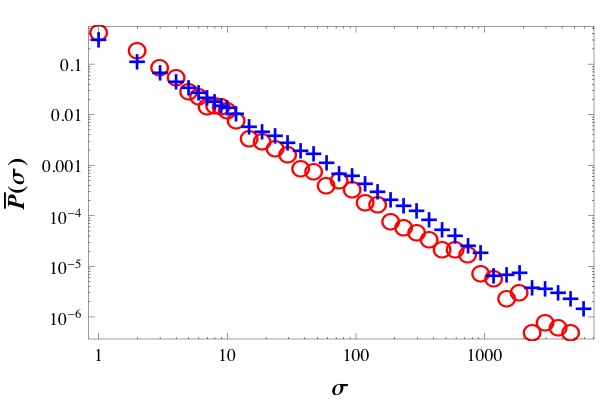} &
\includegraphics[width=50mm,clip=]{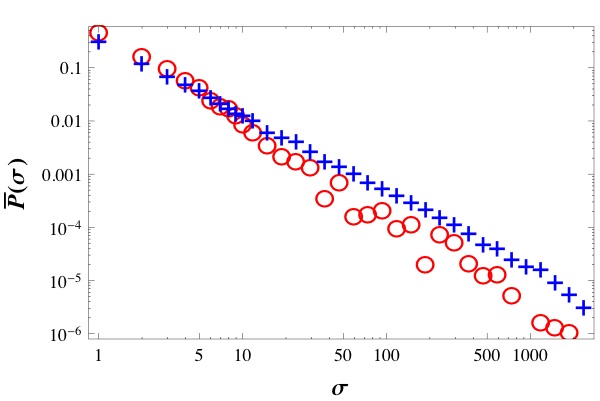}\\
 $\includegraphics[width=50mm,clip=]{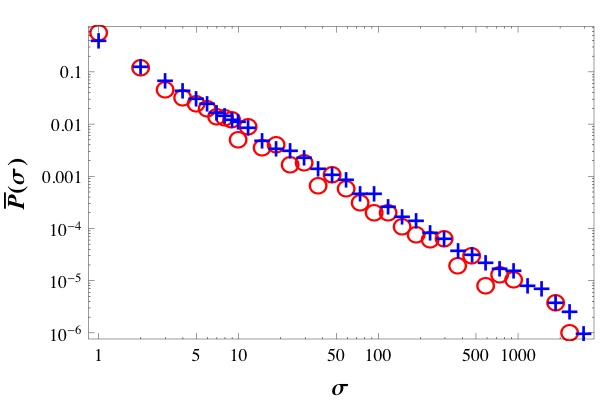}^{\scriptsize{(\star)}}$&
$\includegraphics[width=50mm,clip=]{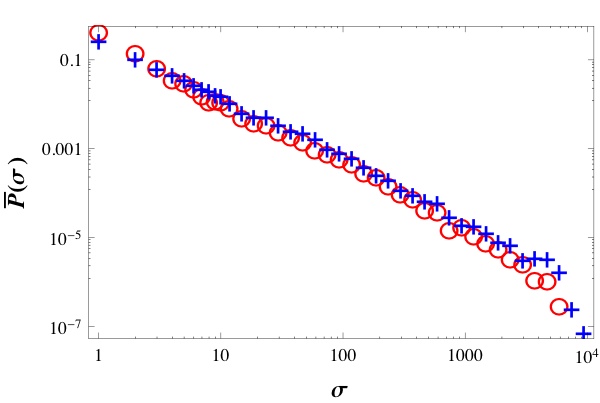}^{\scriptsize{(\star)}} $ &
$\includegraphics[width=50mm,clip=]{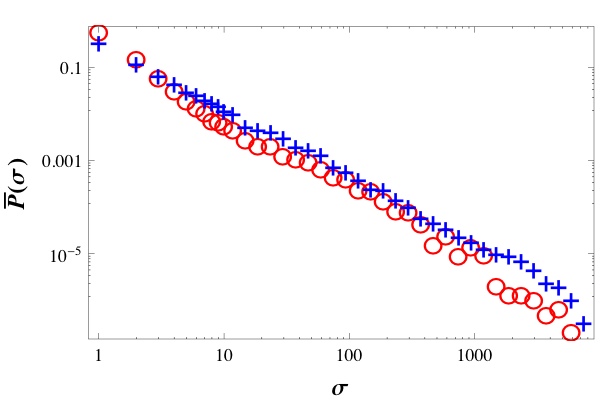}^{\scriptsize{(\star)}} $
\end{tabular}
\caption{\footnotesize Log-log plots of the response-time probability densities $\bar{P}_{R}(\sigma)$ clocked through activity $s$, for a number of typical agents in the long-term email databases DE1 and DE2. The three agents in DE2, whose data extend over periods of five to nine years, are marked by asterisks. Red circles indicate empirical data; blue crosses represent our model predictions.
%
These probability densities are very well fitted by truncated power laws as in eq.~(\ref{truncated_distributions}); the range of individual exponents $\alpha$ in the above distributions goes from 1.293 to 1.845 (see also Table 1 in the main text).
}
\label{RT-TOCK_3/2_supplementary_email}
\end{figure}






\begin{figure}
\begin{center}
\textsc{Response-time distributions of sms clocked through activity $s$}
\end{center}
\hspace*{-20mm}
\begin{tabular}{cccc}
 \includegraphics[width=50mm,clip=]{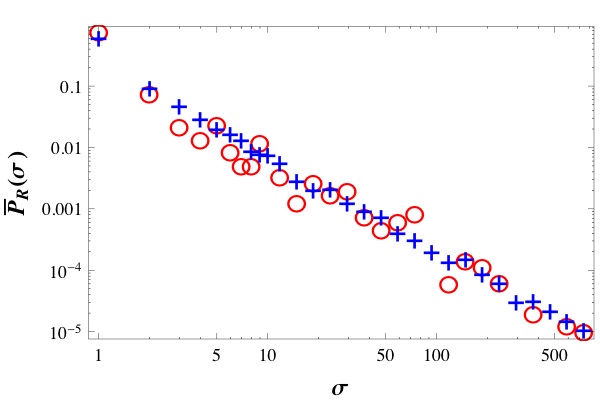} &
\includegraphics[width=50mm,clip=]{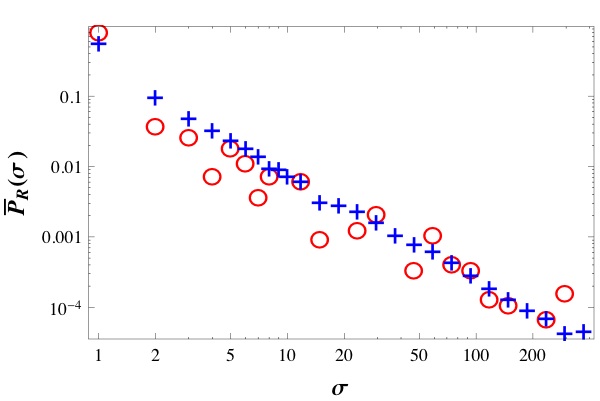} &
\includegraphics[width=50mm,clip=]{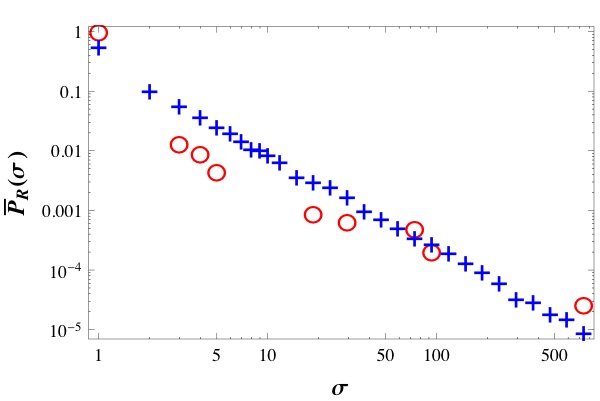}\\
\includegraphics[width=50mm,clip=]{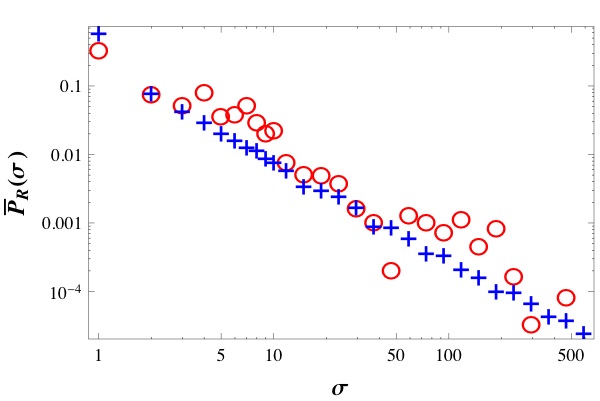}&
\includegraphics[width=50mm,clip=]{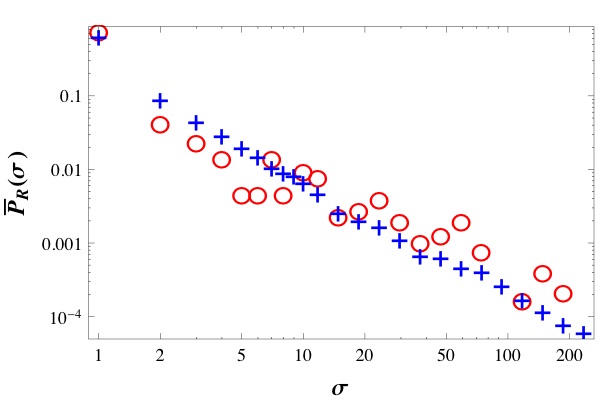} &
\includegraphics[width=50mm,clip=]{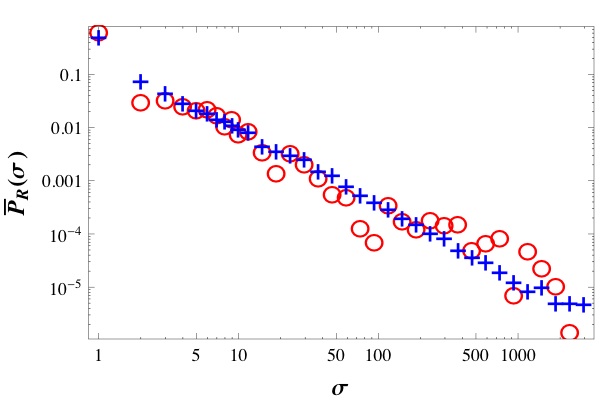}\\
 \includegraphics[width=50mm,clip=]{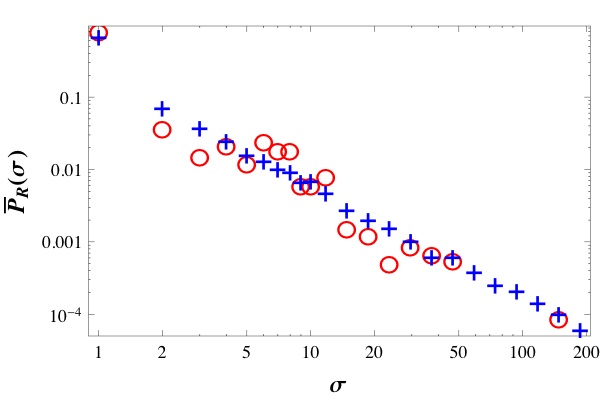} &
\includegraphics[width=50mm,clip=]{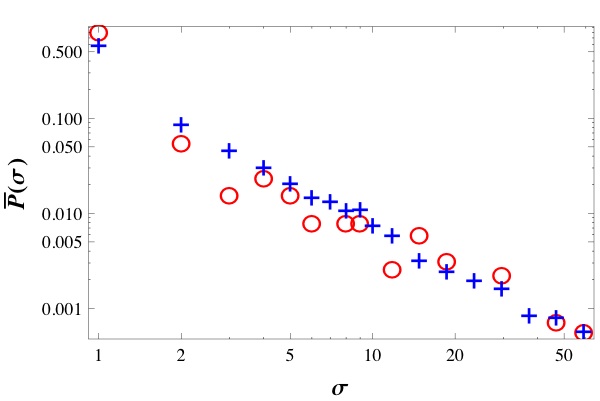}&
\includegraphics[width=50mm,clip=]{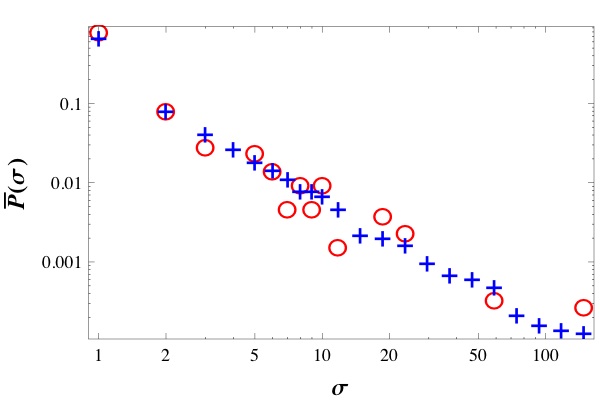} \\
\includegraphics[width=50mm,clip=]{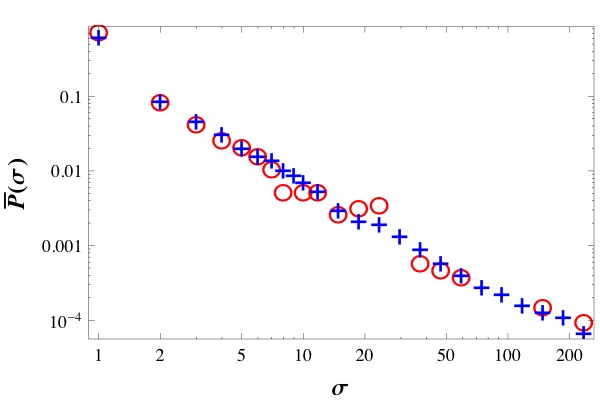}&
 \includegraphics[width=50mm,clip=]{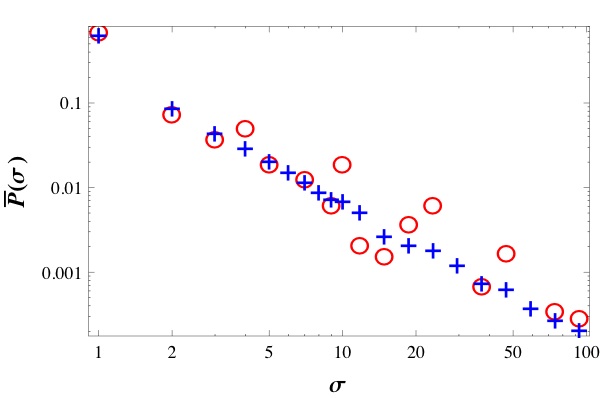} &
\includegraphics[width=50mm,clip=]{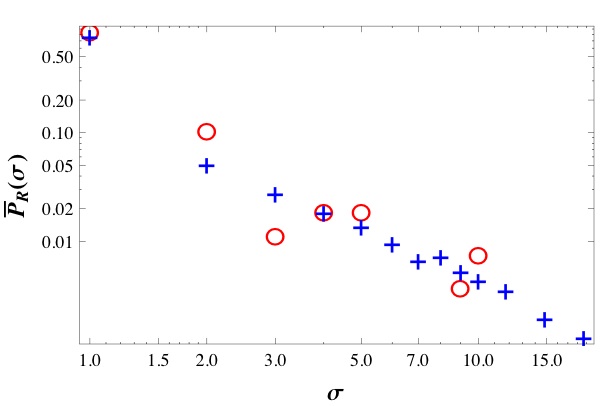}
\end{tabular}
\caption{\footnotesize Log-log plots of the response-time probability densities $\bar{P}_{R}(\sigma)$ clocked through activity $s$, for a number of active agents in the sms database DS1. Red circles indicate empirical data; blue crosses represent our model predictions.
These probability densities can be described by truncated power laws as in eq.~(\ref{truncated_distributions}), with a range of individual exponents $\alpha$ going from 1.160 to 1.668 (see also Table 1 in the main text).
}
\label{RT-TOCK_3/2_supplementary_sms}
\end{figure}


Supporting Figs.~\ref{RT-TOCK_3/2_supplementary_email}-\ref{RT-TOCK_3/2_supplementary_sms} show the empirical RT probability distributions $\bar{P}_{R}(\sigma)$ clocked through activity $s$, for a number of representative agents communicating through email and sms. See also Fig.~2 in the main text for the empirical RT distributions $\bar{P}_{R}(\sigma)$ in letters. For the large majority of the active agents analyzed in all the datasets, such empirical distributions are best fitted by discrete exponentially-truncated power laws
\begin{equation}
	\bar{P}_{R}(\sigma) \sim {\sigma}^{\alpha} e^{-\frac{\sigma}{\lambda}}
\label{truncated_distributions}
\end{equation}
(see eq.~(1) in the main text), in which we have estimated the exponent $\alpha$ and the cutoff parameter $\lambda$ through the maximum likelihood method \cite{newman2009, Rousseau:2000tk}. The log-likelihood ratio test performed on a subset of randomly selected agents shows that other distributions, such as the log-normal, do not reproduce the data with the same accuracy. Table~1 in the main text summarizes the information on the exponents $\alpha$ in (\ref{truncated_distributions}) obtained from the analysis of the different datasets. We find that the distributions $\bar{P}_{R}(\sigma)$ show a remarkable convergence of their empirical exponents to average values close to $-\frac{3}{2}$, across all agents, all databases, and all media. The truncated power-law behavior of the $s$-clocked RT distributions $\bar{P}_{R}(\sigma)$, with average exponent close to $-\frac{3}{2}$ is detected also in the active sms agents, despite the scarcer available statistics.

We have additionally performed systematic Kolmogorov-Smirnov tests for discrete distributions \cite{newman2009, Rousseau:2000tk,Conover:1972vs,Arnold:2011ty} to check that the numerically-simulated $s$-clocked RTs obtained from the queuing model by using the individual inter-arrival times of an agent (see the main text) are compatible with the corresponding empirical data for the same agent. We check compatibility to hold possibly with the exclusion of the durations $\sigma$ smaller or equal to a certain threshold $\sigma_{min}$, according to the procedure suggested in \cite{newman2009}, with values of $\sigma_{min}$ which give compatibility ranges extending from about 2.5 to about 4 orders of magnitude for the various agents. The results of this analysis are reported in Table~\ref{tab:kstests}. We find excellent statistical validation of the compatibility between numerical simulations and empirical data for all the six long-term writers in databases DE2 (long-term email) and DL1 (letters), with $p$-values higher than 0.5 for all six agents.  Also, for database DE1 (two-year email data), we find that more than 75\% of the 300 active writers show $p$-values higher than 0.05, with more than 85\% of them showing $p$-values higher than 0.01 (a common $\sigma_{min}=10$ was chosen for simplicity for all agents in this set).




\begin{table}[htdp]
\caption{\footnotesize Compatibility between numerical simulations and empirical data: $p$-values obtained from KS tests for datasets DE1 (email, two years), DE2 (email, long term), DL1 (letters).}
\begin{center}
\begin{tabular}{cccc}
\hline\hline
\emph{database} &\;\; $\sigma_{min}$\;\; &\;\; $p > 0.05$\;\;  &\;\; $p > 0.01$\;\;  \\ \hline
\textbf{DE1} & 10 & 76.6 \% & 85.6\% \\ \hline \hline 
\end{tabular}
\\
\vspace*{4mm}
\begin{tabular}{ccc}
\hline\hline
\emph{database} &\;\; $\sigma_{min}$\;\; & $p$  \\ \hline\hline
\textbf{DE2} & &  \\ \hline
AL & 2 & 0.62  \\
AP & 3 & 0.61  \\
FC & 3 & 0.55  \\ \hline\hline
\textbf{DL1} & &  \\ \hline
CD & 2 & 0.50  \\
AE & 20 & 0.67   \\
SF & 3 & 0.92  \\ \hline\hline
\end{tabular}
\end{center}
\label{tab:kstests}
\end{table}%




\section{RT statistics clocked through standard time $t$}




\begin{figure}
\begin{center}
\textsc{Response-time distributions for email clocked through time $t$}
\end{center}
\hspace*{-20mm}
\begin{tabular}{cccc}
 \includegraphics[width=50mm,clip=]{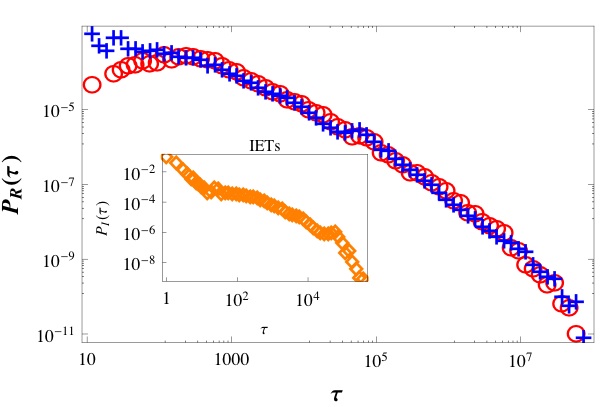} &
\includegraphics[width=50mm,clip=]{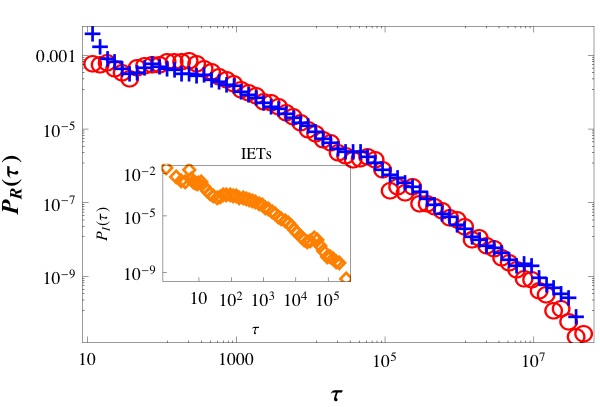} &
\includegraphics[width=50mm,clip=]{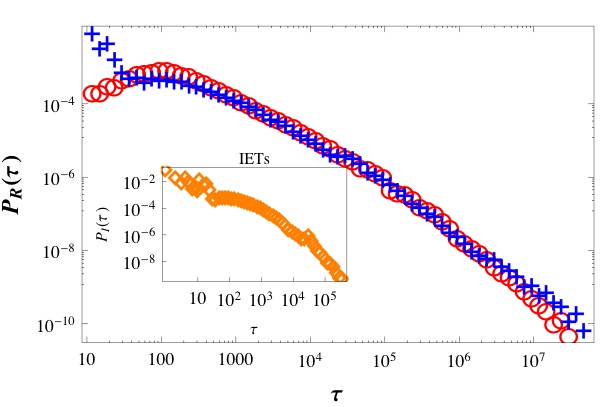} \\
 \includegraphics[width=50mm,clip=]{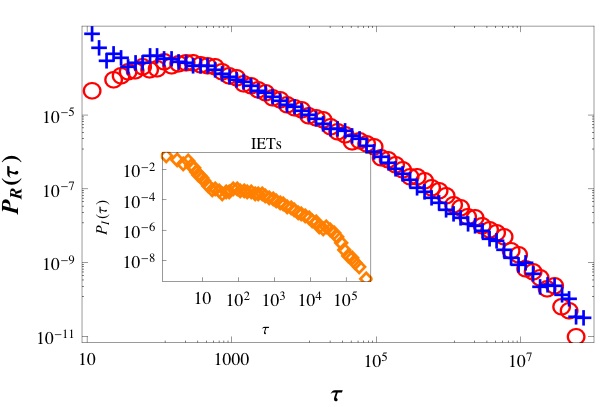} &
\includegraphics[width=50mm,clip=]{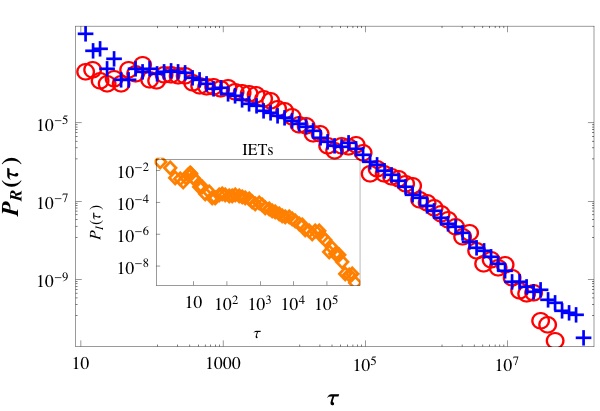} &
\includegraphics[width=50mm,clip=]{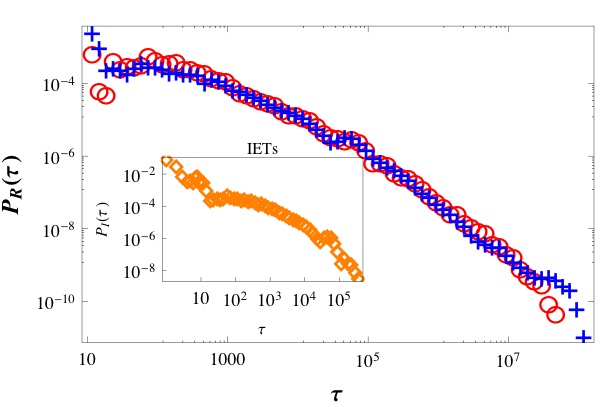} \\
 \includegraphics[width=50mm,clip=]{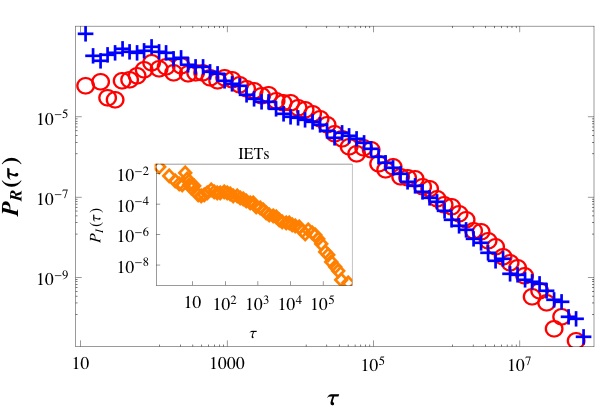} &
\includegraphics[width=50mm,clip=]{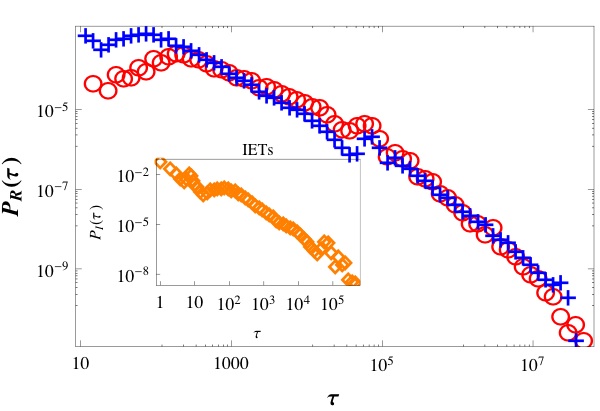} &
\includegraphics[width=50mm,clip=]{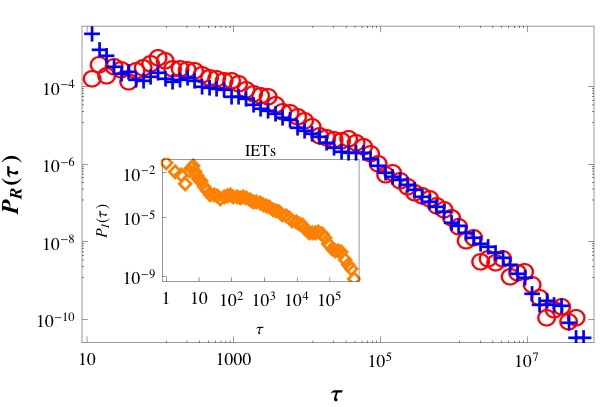} \\
\includegraphics[width=50mm,clip=]{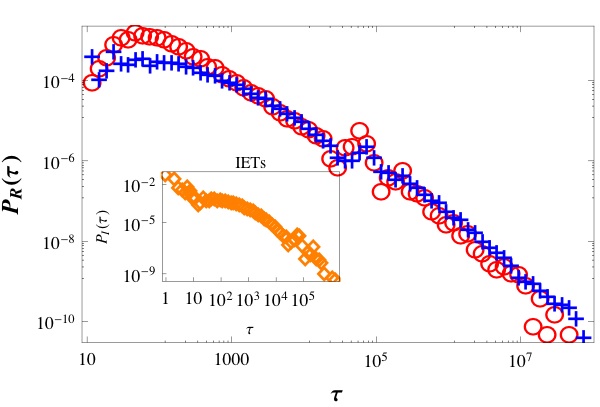}&
 \includegraphics[width=50mm,clip=]{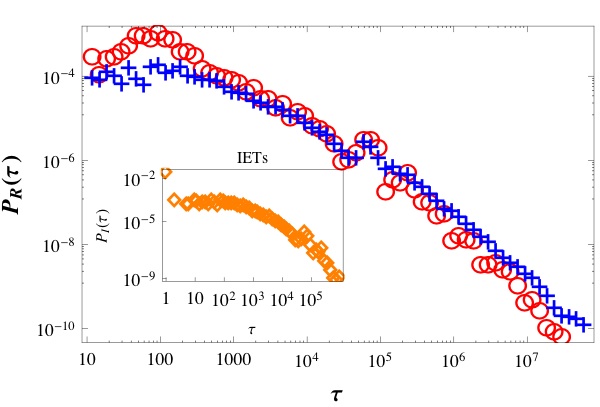} &
\includegraphics[width=50mm,clip=]{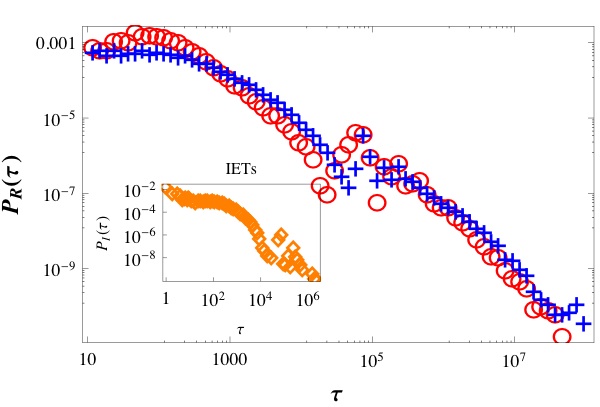}\\
 $\includegraphics[width=50mm,clip=]{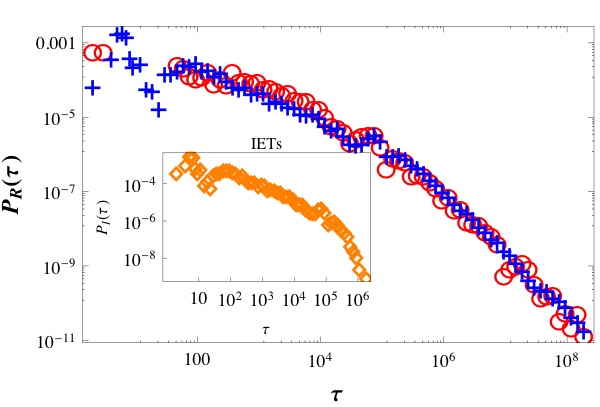}^{\scriptsize{(\star)}} $&
$\includegraphics[width=50mm,clip=]{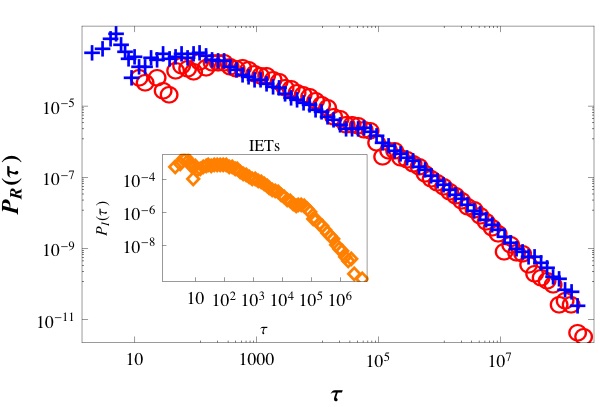}^{\scriptsize{(\star)}} $ &
$\includegraphics[width=50mm,clip=]{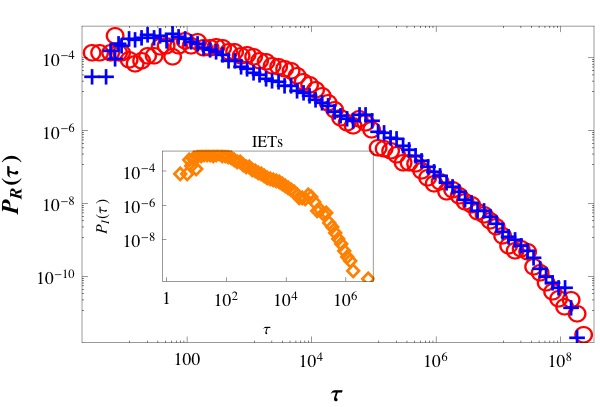}^{\scriptsize{(\star)}} $
\end{tabular}
\caption{\footnotesize Log-log plots of the response-time probability densities $P_{R}(\tau)$ clocked through time $t$ (in seconds), for the same email agents as in Supporting Fig.~\ref{RT-TOCK_3/2_supplementary_email} (agents in the very long term database DE2 are marked by an asterisk). Red circles indicate empirical data; blue crosses represent our model predictions. The insets show the IET distribution $P_{I}(\tau)$ of each agent, used to obtain $P_{R}(\tau)$ from the distribution $\bar{P}_{R}(\sigma)$ in Supporting Fig.~\ref{RT-TOCK_3/2_supplementary_email}, according to eqs.~(\ref{2})-(\ref{1}). These RT distributions for email exhibit bi-modal behavior, with crossover at $\tau \sim T_I$  (see Sect.~\ref{crossover}).
}
\label{tick_bimodal_supplementary_email}
\end{figure}




\begin{figure}
\begin{center}
\textsc{Response-time distributions for sms clocked through time $t$ for sms}
\end{center}
\hspace*{-20mm}
\begin{tabular}{cccc}
 \includegraphics[width=50mm,clip=]{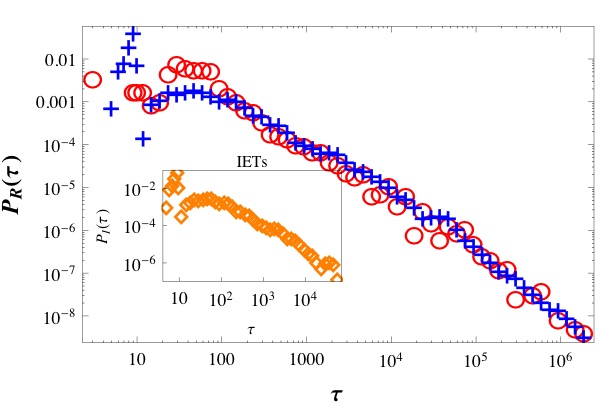} &
\includegraphics[width=50mm,clip=]{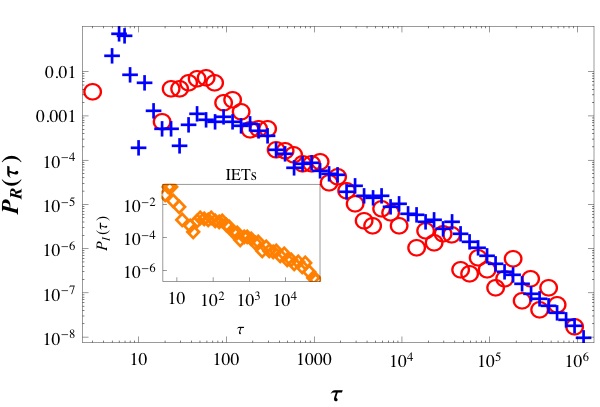} &
\includegraphics[width=50mm,clip=]{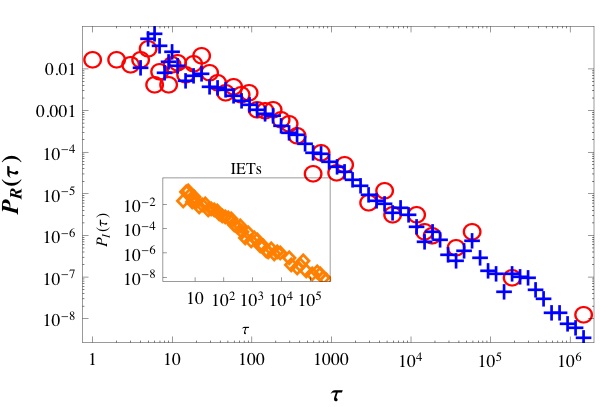} \\
\includegraphics[width=50mm,clip=]{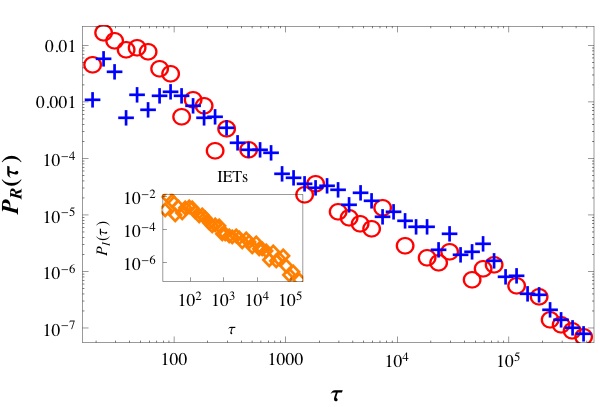} &
\includegraphics[width=50mm,clip=]{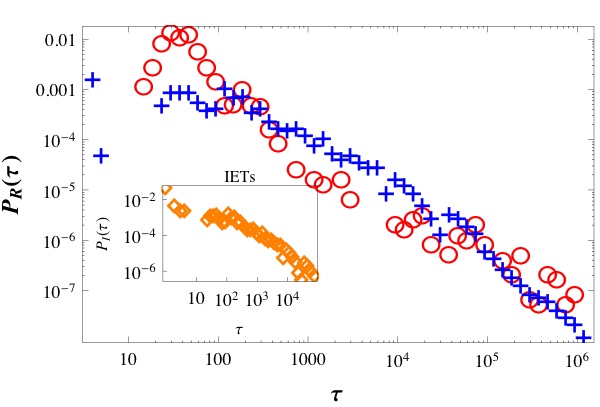} &
\includegraphics[width=50mm,clip=]{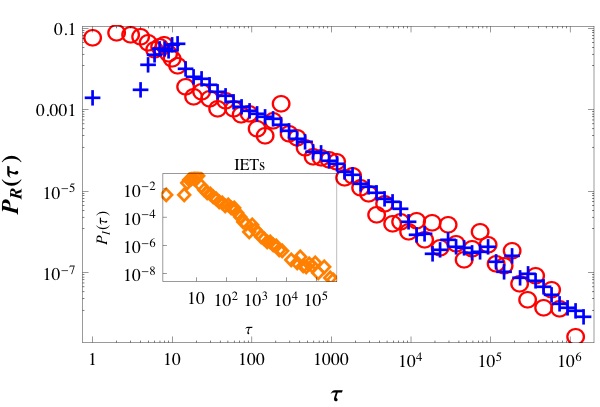} \\
 \includegraphics[width=50mm,clip=]{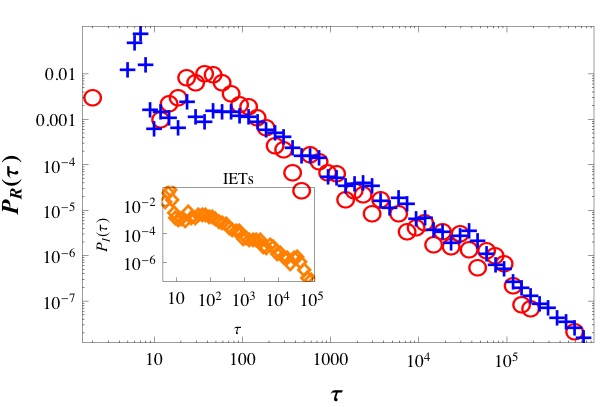} &
\includegraphics[width=50mm,clip=]{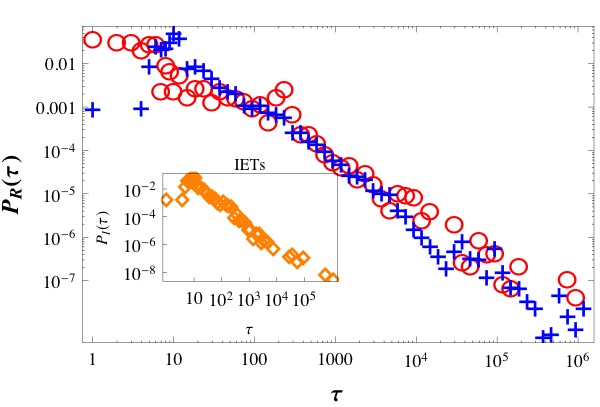} &
\includegraphics[width=50mm,clip=]{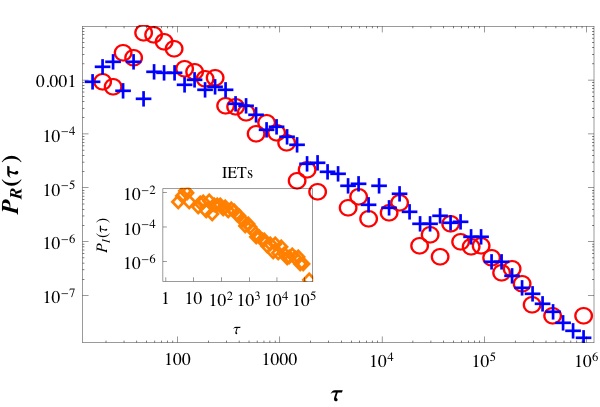} \\
\includegraphics[width=50mm,clip=]{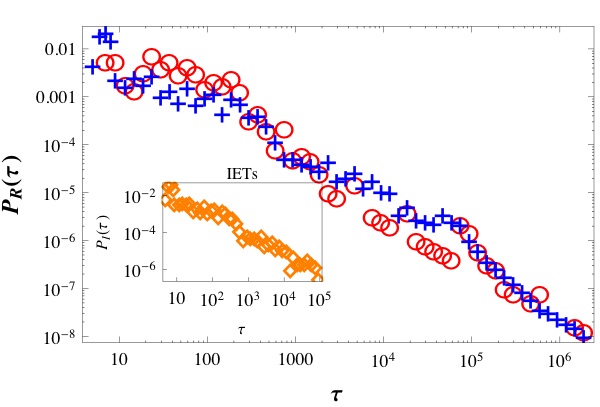}&
 \includegraphics[width=50mm,clip=]{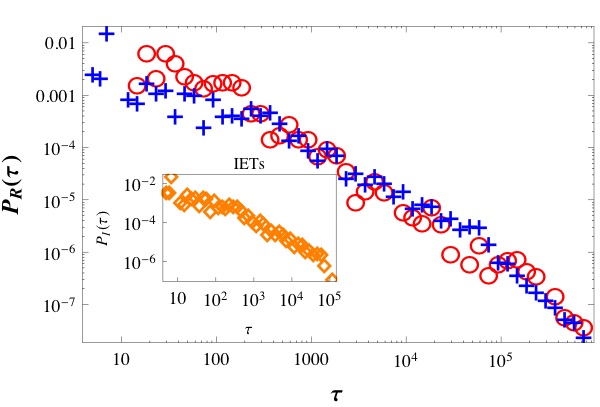} &
\includegraphics[width=50mm,clip=]{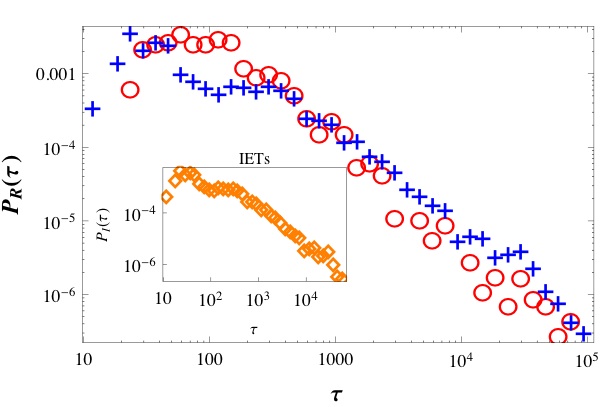}

\end{tabular}
\caption{\footnotesize Log-log plots of the response-time probability densities $P_{R}(\tau)$ clocked through time $t$ (in seconds), for the same sms agents as in Supporting Fig.~\ref{RT-TOCK_3/2_supplementary_sms}. Red circles indicate empirical data; blue crosses represent our model predictions. The insets show the IET distribution $P_{I}(\tau)$ of each agent, used to obtain $P_{R}(\tau)$ from the distribution $\bar{P}_{R}(\sigma)$ in Supporting Fig.~\ref{RT-TOCK_3/2_supplementary_sms}, according to eqs.~(\ref{2})-(\ref{1}). The bi-modal behavior in these RT distributions, with crossover at $\tau \sim T_I$ (see Sect.~\ref{crossover}), is discernible despite the short time window of the sms database.
}
\label{tick_bimodal_supplementary_sms}
\end{figure}


Supporting Figs.~\ref{tick_bimodal_supplementary_email}-\ref{tick_bimodal_supplementary_sms} show the empirical RT probability distributions $P_{R}(\tau)$ clocked through $t$, relative to the same representative agents as in Supporting Figs.~\ref{RT-TOCK_3/2_supplementary_email}-\ref{RT-TOCK_3/2_supplementary_sms}, for email and sms (see Fig.~3 in the main text for the empirical distributions $P_{R}(\tau)$ in letters).

Given the IET distribution $P_{I}(\tau)$ of an agent $\mathcal{A}$, the relation between the two RT distributions $P_{R}(\tau)$ and $\bar{P}_{R}(\sigma)$ of $\mathcal{A}$ can be obtained in a natural way as follows. We consider independent random variables $N$ and $\rho_{I}(h)$, $h=1,2,\dots$, where: (\emph{i}) $N$ represents the number of activities between a message received and the response to it, sampled from the RT distribution $\bar{P}_{R}(\sigma)$; and (\emph{ii}) the $\rho_{I}(h)$ are all identically-distributed, sampled from the IET distribution $P_{I}(\tau)$. Then the corresponding time-clocked RT distribution, obtained by separating any two consecutive activities of $\mathcal{A}$ by an inter-event time $\rho_{I}(h)$, is described by the compounding process:
\begin{equation}
\label{2}
\rho_R =\sum_{h=1}^N\rho_{I}(h)
\end{equation}
taking values in $\mathbb{N}^+$, whose probability density $P_{R}(\tau)$, computed by conditioning, is given by:
\begin{equation}
\label{1}
P_{R}(\tau) =\sum_{\sigma\geq1}\mbox{Prob}\left(\sum_{h=1}^\sigma\rho_{I}(h)=\tau\right)\bar{P}_{R}(\sigma).
\end{equation}
Numerical simulations confirm the above relations hold for the empirical distributions $P_{R}(\tau)$, $P_{I}(\tau)$, and $\bar{P}_{R}(\sigma)$, in all media. Notice that the $s$- vs. $t$-clocked RT statistics $\bar{P}_{R}(\sigma)$ and $P_{R}(\tau)$ of an agent can have significantly different number of filled bins, due to the role of the IET statistics $P_{I}(\tau)$ in each bin of $\bar{P}_{R}(\sigma)$, according to (\ref{2})-(\ref{1}); this effect is particularly evident for instance in the RT statistics of the last sms agent in Supporting Figs.~\ref{RT-TOCK_3/2_supplementary_sms} and \ref{tick_bimodal_supplementary_sms}. 

We show in Sect.~\ref{bimodal} that distributions $P_{R}(\tau)$ satisfying (\ref{2})-(\ref{1}) exhibit bi-modal behavior with crossover at $\tau \sim T_I$, where $T_{I}  \sim \frac{<\tau^2>}{<\tau>}$ is the characteristic time of the IET distribution $P_{I}(\tau)$. The bi-modality in the empirical $P_{R}(\tau)$ is particularly evident in the email RT distributions in Supporting Fig.~\ref{tick_bimodal_supplementary_email}, which involve the highest number of decades among all media, from seconds to several years (see the discussion in Sect.~\ref{crossover}).



\section{Bi-modality and crossover in the $t$-clocked RT distributions $P_{R}(\tau)$\label{bimodal}}




\subsection{An analysis of the bi-modality of the RT distribution $P_{R}(\tau)$}


We study the main features of the $t$-clocked RT distributions $P_{R}(\tau)$ obtained from (\ref{2})-(\ref{1}), assuming%
\footnote{~These simplified forms for the $s$-clocked RT distribution $\bar{P}_{R}(\sigma)$ and the IET distribution $P_{I}(\tau)$ are justified because there is in general a separation of scales in the cutoffs of $P_{I}(\tau)$ vs.~$P_{R}(\tau)$, the latter being much larger, which allows us to assume $\bar{P}_{R}(\sigma)$ in (\ref{truncated_distributions}) to have, for the present analysis, an effectively infinite cutoff. Likewise, for our purposes the heavy-tailed empirical IET distribution $P_{I}(\tau)$ can roughly be approximated by an exponentially truncated power law with exponent $\beta$ and characteristic time $T_{I}$, i.e.~$P_{I}(\tau) \sim {\tau}^{\beta}\mbox{exp}(-\tau/T_{I})$, where $\beta \sim -1$ for email, see \cite{Eckmann:2004vn, Barabasi:2005yq,Vazquez:2006zr,Malmgren:2008ys}, while the IET tails decrease much faster for the other media.
}
 $\bar{P}_{R}(\sigma) \sim \sigma^{\alpha}$, with $\alpha$ near $-\frac{3}{2}$, and $P_{I}(\tau) \sim {\tau}^{\beta}\mbox{exp}(-\tau/T_{I})$. From this, the distribution $P_{R}(\tau)$ in (\ref{2})-(\ref{1})  has a bi-modal character, which can be understood for instance by computing the Generating Function (GF) \cite{combinatorics} of the random variable $\rho_R$ in (\ref{2}), which is defined as
\begin{equation}
\label{gf}
G_{\rho_R}(z)=\sum_{\tau\geq 1} P_R(\rho_R=\tau) z^{\tau}, \quad z \in [0, 1];
\end{equation}
the GF encodes the law of $\rho_R$, as $P_R(\rho_R=\tau)=\frac{1}{{\tau}!}\frac{\text{d}^{\rho} G_{\tau_R}}{\text{d} z^{\tau}}(0)$.

A standard computation \cite{combinatorics} shows that for the process \eqref{2} one obtains $G_{\rho_R}=G_{N}\circ G_{\rho_I}$. Considering for definiteness the case $\alpha= -\frac{3}{2}$ and $\beta=-1$, we have:
\begin{equation}
\label{gf2}
G_{\rho_R}(z)=\frac{1}{\zeta(3/2)}Li_{3/2}\left(\frac{\log(1-e^{-1/T_{I}}z)}{\log(1-e^{-1/T_{I}})}\right),
\end{equation}
where $Li_{w}$ is the polylogarithm of complex order $w$, and $\zeta$ is the Riemann zeta function.

For any given $T_{I}$, we can derive the asymptotic behavior of the probability $P_{R}(\tau)$ for large $\tau$ through a singularity analysis on the GF $G_{\rho_R}$ (see \cite{combinatorics}), because the behavior of the GF near its lowest-norm singularity identifies how $P_{R}(\tau)$ decays for large $\tau$. In particular, let $\xi\geq 1$ be the GF's lowest-norm singularity, and let $(1-z/\xi)^{-\gamma}$, $\gamma\in\mathbb{R}\setminus\mathbb{Z}_{\leq 0}$, be the leading term in the expansion of the GF near $\xi$. This implies that the asymptotic behavior of the GF is $C\xi^{-t}t^{\gamma-1}$.
In our case, since $Li_{3/2}$ is singular on the reals $\geq1$, and is analytic with $G_{\rho_I}(1)=1$, the GF $G_{\rho_R}$ indeed has a lowest-norm singularity $\xi=1$, whose type is determined by the expansion of $Li_{3/2}$ near 1, in which the leading exponent of $(1-z)$ is $-\frac{1}{2}$ (see the definition of supercritical composition and Theorem VI.7 in \cite{combinatorics}). Then we conclude, as mentioned above, that asymptotically $P_{R}(\tau)$ decays in the same way as $\bar{P}_{R}(\sigma)$, i.e.~as a power law with exponent $-\frac{3}{2}$.
%

On the other hand, to see the behavior for $\tau \ll T_I$, we consider $T_{I}$ large. In this case
\begin{equation}
\label{5}
\frac{\text{d}^{\tau}  G_{\rho_R}}{\text{d} z^{\tau}}(0)\sim \frac{1}{-\log(T_I) \zeta(3/2)}\frac{\text{d}^{\tau} \log(1-z)}{\text{d} z^{\tau}}(0),
\end{equation}
so that $P_R(\tau)={\tau}^{-1}$ for $T_{I}\rightarrow\infty$. Thus, the larger $T_{I}$, the closer the distribution $P_R(\tau)$ follows a power-law behavior with exponent $\beta = -1$, i.e.~$P_R(\tau) \sim P_I(\tau)$, for $\tau \ll T_I$.
%



%
\begin{figure}  
\centering
\includegraphics[width=100mm,clip=]{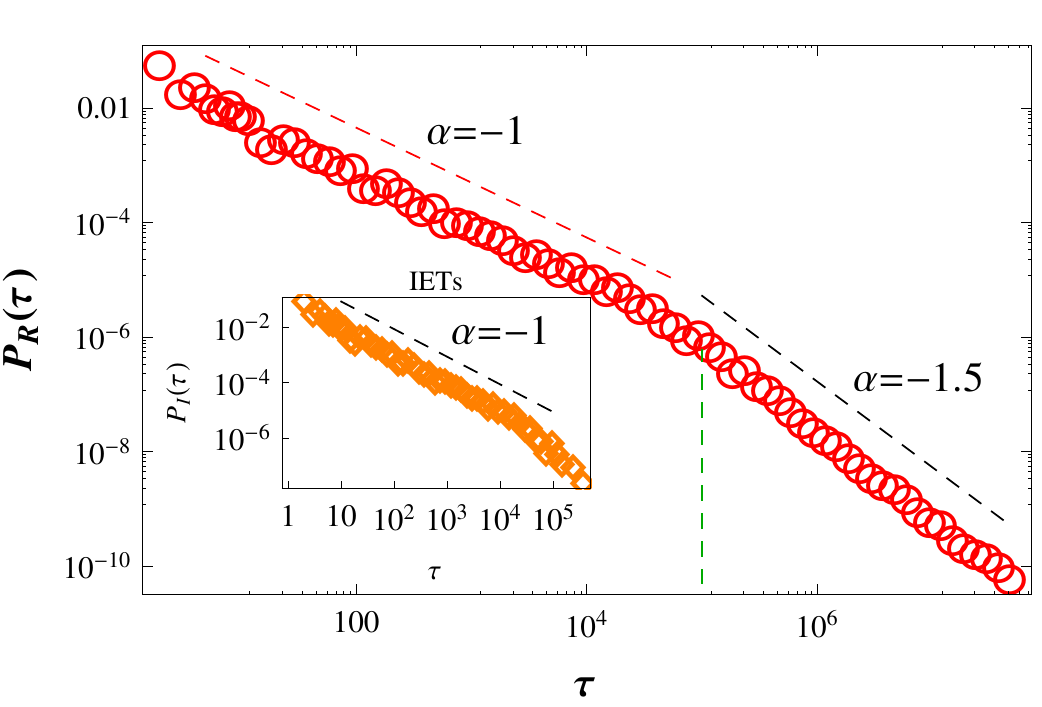}
\caption{\footnotesize  Log-log plot of the bi-modal $t$-clocked RT distribution $P_{R}(\tau)$, computed through eqs.~(\ref{2})-(\ref{1}), from the $s$-clocked RT distribution $\bar{P}_{R}(\sigma) \sim {\sigma}^{-3/2}$, and the IET distribution $P_{I}(\tau) \sim t^{-1}\mbox{exp}(-t/T_{I})$ shown in the inset. According to predictions, the tail of $P_{R}(\tau)$ for large $\tau$ follows a power law with the same exponent $-\frac{3}{2}$ as $\bar{P}_{R}(\sigma)$, while for small $\tau$, $P_{R}(\tau)$ follows the features of the IET distribution $P_{I}(\tau)$ (in this case $P_{R}(\tau)$ initially follows a power law with exponent close to $-1$). The crossover (green dashed vertical line) occurs for $\tau \sim T_{I}$, where $T_{I} \sim \frac{<\tau^2>}{<\tau>}$ is the characteristic time of $P_{I}(\tau)$, with value $T_I \sim 10^5$ in the present case. The straight dashed lines, drawn to guide the eye, have the indicated exponents $-\frac{3}{2}$ and $-1$.
}
\label{ginocchio}
\end{figure}



The two distinct regimes, for small $\tau$ vs.~large $\tau$, that characterize the distribution $P_R(\tau)$, are explicitly shown in Supporting Fig.~\ref{ginocchio}, where the relations (\ref{2})-(\ref{1}) are simulated numerically for $\bar{P}_{R}(\sigma) = \sigma^{-3/2}$, and $P_{I}(\tau) = {\tau}^{-1}\mbox{exp}(-\tau/T_{I})$, with $T_{I} \sim 10^5$. Supporting Fig.~\ref{ginocchio} shows explicitly the bi-modality of the resulting distribution $P_R(\tau)$, indicating also that the scaling crossover indeed occurs for $\tau \sim T_{I}$.  Further mathematical analysis and numerical simulations show that such bi-modality is a stable feature of the compound distribution $P_R(\tau)$ in (\ref{2})-(\ref{1}), the two distinct regimes being identifiable also for a range of exponents $\alpha \sim -\frac{3}{2}$, and $\beta \leq -1$. Precisely, we find that: (\emph{a}) for large $\tau$ the distribution $P_R(\tau)$ in (\ref{2})-(\ref{1}) always shows tails with the same scaling exponent as $\bar{P}_R(\sigma) \sim \sigma^{\alpha}$ for $\alpha$ near $-\frac{3}{2}$; furthermore (\emph{b}) for small $\tau \ll T_I$, $P_R(\tau)$ is influenced by the scaling features of $P_{I}(\tau)$ when $\beta$ grows smaller than $-1$, in which case the initial (i.e.~small $\tau$) scaling exponent of $P_R(\tau)$ tends to grow closer to the exponent $\alpha$ of the scaling tails of $P_{R}(\tau)$ (for instance, for fixed $\alpha = -\frac{3}{2}$, as $\beta$ grows smaller than $-1$ in $P_I(\tau)$, the scaling exponent of $P_{R}(\tau)$ for small $\tau$ decreases, going from $-1$ towards $-\frac{3}{2}$).

Even more relevant for the analysis of human correspondence, the simulations show that the bi-modal behavior of $P_{R}(\tau)$ (with $P_{R}(\tau)$ reflecting the features $P_{I}(\tau)$ for $\tau \ll T_{I}$, and  crossover for large $\tau$ to a scaling tail with the same exponent $\alpha$ as $\bar{P}_R(\sigma)$) is observed when the IET distribution $P_{I}(\tau)$ departs significantly from a truncated power law, but admits heavy tails with a finite characteristic time $T_{I} \sim \frac{<\tau^2>}{<\tau>}$, as is the case for the empirical IET distributions $P_{I}(\tau)$ for all media (letters, email, sms), see Supporting Figs.~\ref{tick_bimodal_supplementary_email}-\ref{tick_bimodal_supplementary_sms} and Fig.~3 in the main text. Typical empirical values of $T_I$ are given below.


\subsection{Bi-modality in the empirical RT distributions $P_{R}(\tau)$\label{crossover}}



The above analysis of the bi-modality of the distribution $P_{R}(\tau)$ in (\ref{2})-(\ref{1}) indicates the origin of the complex, media-dependent, features observed in the empirical RT probabilities $P_{R}(\tau)$. We see in Supporting Figs.~\ref{tick_bimodal_supplementary_email}-\ref{tick_bimodal_supplementary_sms}, and Fig.~3 in the main text, that the tails of the empirical $P_{R}(\tau)$ scale with the same exponents exponents $\alpha$ as the corresponding $s$-clocked distributions $\bar{P}_{I}(\sigma)$, while for small $\tau$ the empirical $P_{R}(\tau)$ are affected by the features of the corresponding empirical IET distributions $P_{I}(\tau)$, which are different in the three media, reflecting specific ways and styles in which the three technologies are used in written communication.





\begin{figure}[t]
\begin{center}
\includegraphics[width=80mm]{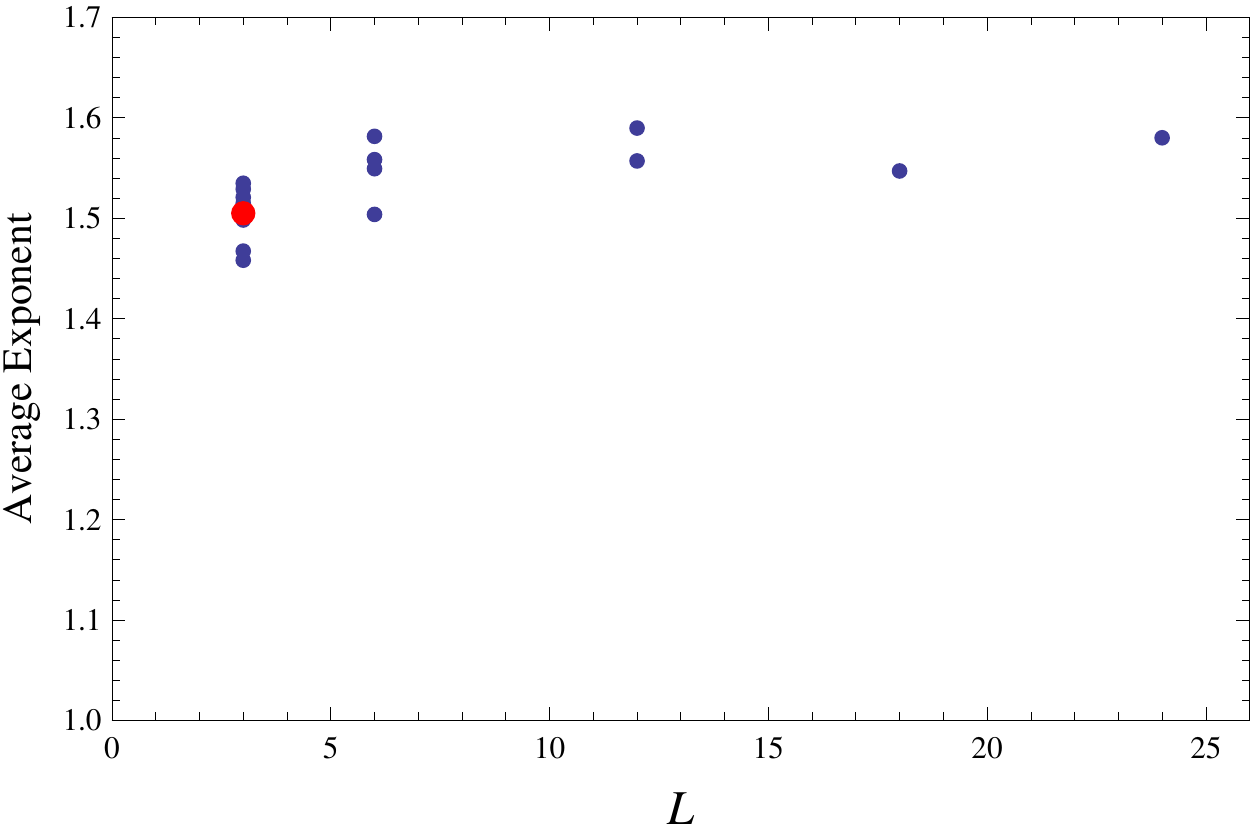}
\caption{\footnotesize Average value of the individual exponents for the $s$-clocked RT probabilities $\bar{P}_{R}(\sigma)$ in (\ref{truncated_distributions}) referred to all the sampled time windows, with durations of three, six, twelve, and eighteen months, extracted from the two-year email database DE1 ($L$ denotes the window length in months). The red dot indicates the average value of the distribution of individual exponents for the three-months email database DE3 from \cite{Eckmann:2004vn}.
}
\label{exponents_windows}
\end{center}
\end{figure}


In detail, we observe that:

(\emph{i})~~The empirical distributions $P_{R}(\tau)$ for email agents in the long-term databases DE1 and DE2 show the clearest bi-modality, with a crossover at $\tau \sim T_{I}$ (where $T_{I} \sim 10^5$ sec is the characteristic time of typical IET distributions $P_{I}(\tau)$ for email). This is because (see points $(a)$-$(b)$ above) the crossover separates a scaling tail with exponent $-\frac{3}{2}$ for large $\tau$, from a small-$\tau$ regime reflecting the features of $P_{I}(\tau)$ (thus having an approximated scaling exponent $\beta \sim -1$, see~\cite{Eckmann:2004vn,Malmgren:2008ys}). See Supporting Fig.~\ref{tick_bimodal_supplementary_email} and Fig.~3 in the main text.

(\emph{ii})~~The bi-modality of $P_{R}(\tau)$ is much less clear for sms, because the typical empirical IET distributions $P_{I}(\tau)$ have heavy tails decreasing much faster than email: this means that the corresponding RT distributions $P_{R}(\tau)$ behave, for small $\tau$, rather similarly to their own large-$\tau$ tails (see point $(b)$ above). The crossover at $\tau \sim T_{I}$ is still somewhat distinguishable in these bi-modal RT distributions, where $T_{I} \sim 10^4$ sec is a typical value of the IET characteristic time for sms (see Supporting Figs.~\ref{tick_bimodal_supplementary_sms} and Fig.~3 in the main text).

(\emph{iii})~~The same discussion as in (\emph{ii}) holds for the RT distribution $P_{R}(\tau)$ of letters, but in this case with a small $T_{I} \sim 5$ days ($T_{I} \sim 10$ days for S.~Freud), so that the scaling tails of $P_{R}(\tau)$ are largely predominant, $P_{R}(\tau)$ showing a barely discernible small-$\tau$ regime (see panel (a) of Fig.~3 in the main text).

\begin{figure}
\begin{center}
\begin{tabular}{cc}
\includegraphics[width=65mm,clip=]{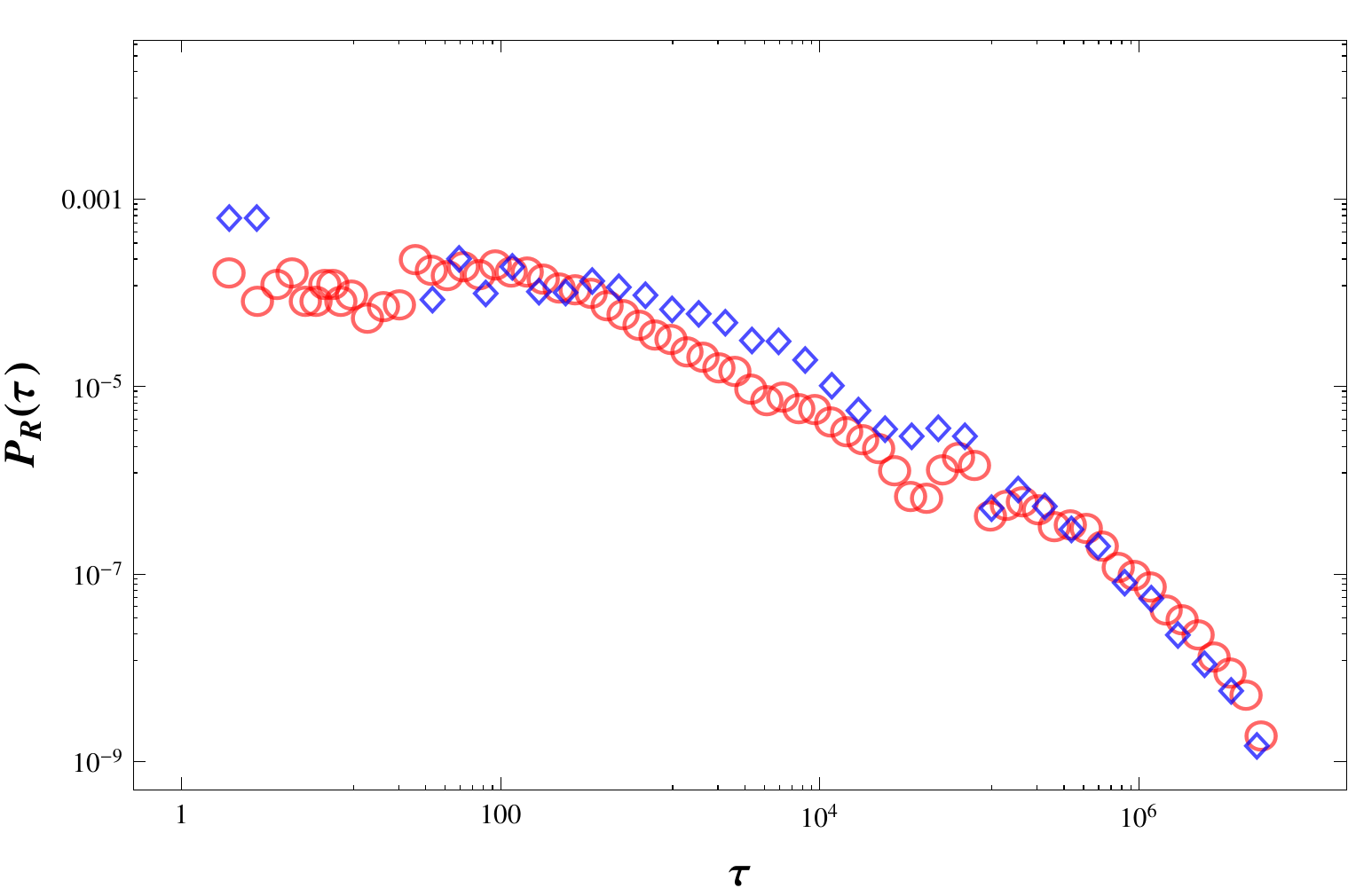} \qquad \qquad &
\includegraphics[width=65mm,clip=]{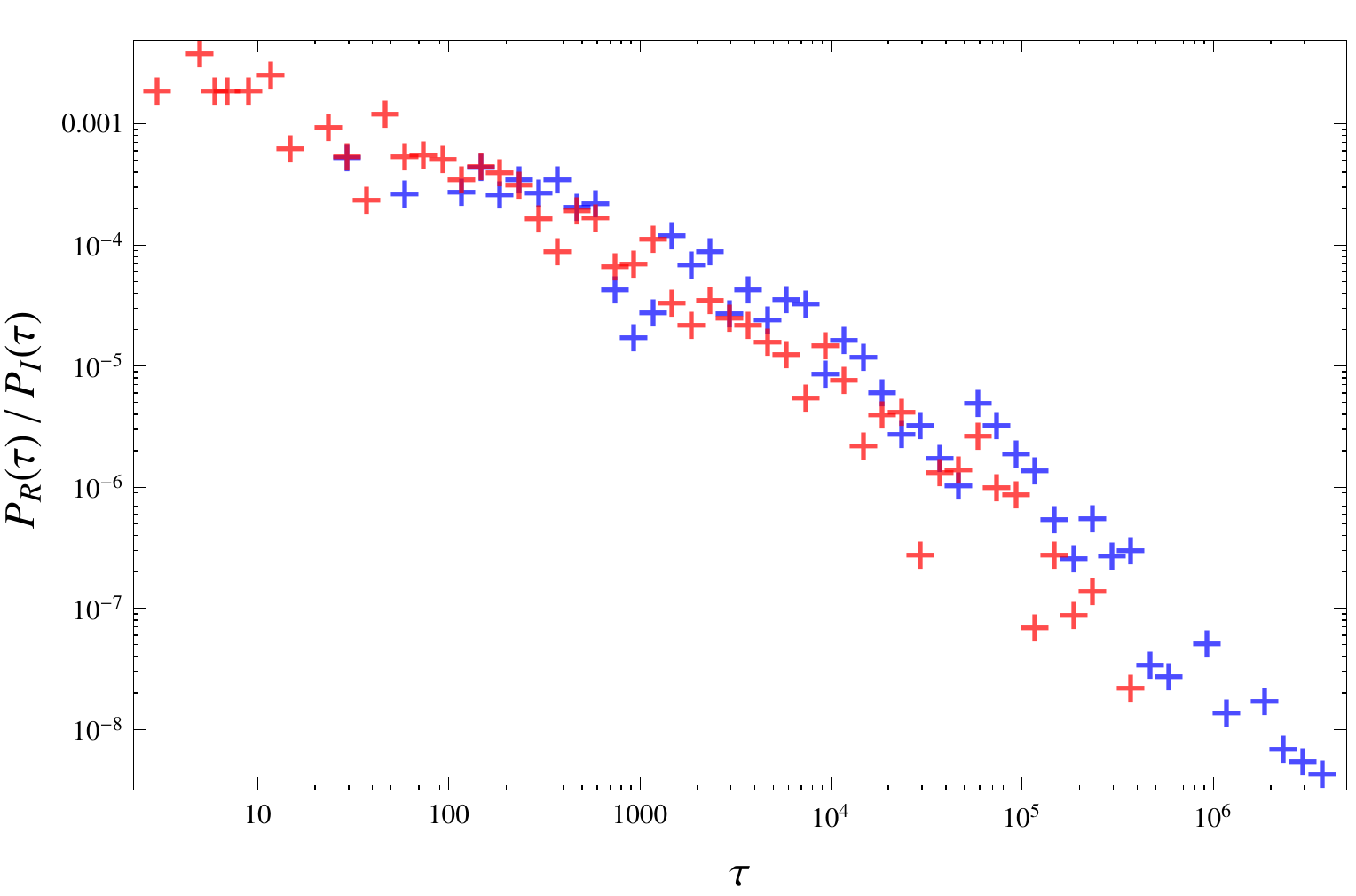}\\
(a) &(b)\\
\end{tabular}
\end{center}
\caption{\footnotesize  (a) Log-log plot of the collective $t$-clocked RT distribution $P_{R}(\tau)$ for 31 three-month windows sampled from the seven-year email data of an user in database DE2 (blue symbols), compared to the collective RT distribution $P_{R}(\tau)$ computed by aggregating the email data of the agents in the three-month database DE3 (red symbols). The two distributions are superposable. (b) Log-log plot of the RT distribution $P_{R}(\tau)$ (blue symbols) and the IET distribution $P_{I}(\tau)$ (red symbols) for a typical email agent in database DE3. Due to the short observation window in this dataset, the two distributions show a strong correlation, see point (\emph{iv}) in Sect.~\ref{crossover}.
}
\label{three_month_windows_supplementary}
\end{figure}

(\emph{iv})~~When the cutoffs in the observed the RTs and IETs are not well separated in scale, the small-$\tau$ regime predominates in the RT distribution $P_{R}(\tau)$; in this case the scaling regime at the tail of the  $P_{R}(\tau)$ can be confused with the cutoff, and the small-$\tau$ regime is predominant in $P_{R}(\tau)$. The RTs thus result to be correlated to the IETs, with $P_{R}(\tau)$ showing a behavior qualitatively similar to $P_{I}(\tau)$, possibly except for its extreme tail. This happens for instance in the empirical distributions $P_{R}(\tau)$ of agents in the earlier email database DE3, as a consequence of the short (three-month) observation window. A typical example of this effect in DE3 is shown in panel (b) of Supporting Fig.~\ref{three_month_windows_supplementary}. The correlation of RTs and IETs had also been discussed in~\cite{Vazquez:2006zr}, based on different reasons than presently proposed.

These points clarify how the complex interplay of the distributions $P_{I}(\tau)$ and $\bar{P}_{R}(\sigma)$, and of their cutoffs, generate the specific bi-modal features of the $t$-clocked RT probabilities $P_{R}(\tau)$ observed empirically in the different media. This contributes to explain the origin of the earlier controversial judgements made in the literature regarding these time statistics of written communication, especially in email \cite{Barabasi:2005yq,Vazquez:2006zr,Eckmann:2004vn,JohansenCommento2006,Oliveira:2005fk,quwang2011,malmgremCritica2006}, as the RTs have neither power-law nor log-normal behavior.



\section{Comparison of the new long-term \\ with the earlier short-term data for email\label{windows}}






We check for the internal consistency of the results on the scaling exponents in the two-year email database DE1, and for cross-validation of this from the data in the independently-collected three-month email dataset DE3. We have computed the exponents $\alpha$ for the $s$-clocked distributions $\bar{P}_{R}(\sigma)$ in (\ref{truncated_distributions}) obtained by sampling the two-year data in DE1 through consecutive three-, six-, twelve- and eighteen-month windows.
Supporting Fig.~\ref{exponents_windows} summarizes the average exponent values obtained in this way, which are all consistent with each other, and close to $-\frac{3}{2}$, across all windows lengths within database DE1, as well as across the databases DE1 and DE3 for the three-month windows. 
%




A further check on the compatibility of the new empirical data on long-term email use with the earlier short-term email data DE3, is obtained by sampling a number of randomly selected three-month windows from the seven-year data email belonging to an agent in database DE2. Supporting Fig.~\ref{three_month_windows_supplementary}, panel (a), shows that the collective RT distribution $P_{R}(\tau)$ obtained from the aggregate three-month window data is superposable to the collective RT distribution obtained from the aggregate agent data in the three-month database DE3.

\bibliographystyle{unsrt}
\bibliography{emails.bib}

\begin{thebibliography}{10}

\bibitem{nakamura2008}
Toru Nakamura, Toru Takumi, Atsuko Takano, Naoko Aoyagi, Kazuhiro Yoshiuchi,
  Zbigniew~R Struzik, and Yoshiharu Yamamoto.
\newblock Of mice and men---universality and breakdown of behavioral
  organization.
\newblock {\em PLoS One}, 3(4):e2050, 2008.

\bibitem{HANAI:kx}
K~Hanai, M~Ozaki, D~Yamauchi, and Y~Nakatomi.
\newblock {Scale Free Dynamics Involved in the Ant Locomotion}.
\newblock {\em Proceedings of the 2006 WSEAS Int. Conf. on Cellular and
  Molecular Biology, Biophysics and Bioengineering}, 2006.

\bibitem{Henderson:2001yq}
T.~Henderson and S.~Bhatti.
\newblock {Modelling user behaviour in networked games}.
\newblock {\em MULTIMEDIA '01: Proc. of the 9th ACM international conference on
  Multimedia}, pages 90--94, 2001.

\bibitem{Wang:2010vn}
Qing Wang and Jin-Li Guo.
\newblock {Human dynamics scaling characteristics for aerial inbound logistics
  operation}.
\newblock {\em Physica A-Statistical Mechanics And Its Applications},
  389(10):2127--2133, 2010.

\bibitem{wang2012random}
Chunyan Wang and Bernardo~A Huberman.
\newblock How random are online social interactions?
\newblock {\em Scientific reports}, 2, 2012.

\bibitem{Dezso:2006kx}
Z~Dezso, E~Almaas, A~Luk{\'a}cs, B~R{\'a}cz, I~Szakad{\'a}t, and A~L
  Barab{\'a}si.
\newblock {Dynamics of information access on the web}.
\newblock {\em Phys. Rev. E}, 73:066132, June 2006.

\bibitem{Goncalves:2008yq}
Bruno Gon{\c c}alves and Jose~J. Ramasco.
\newblock {Human dynamics revealed through Web analytics}.
\newblock {\em Phys. Rev. E}, 78:026123, 2008.

\bibitem{Gao:2010fj}
Lei Gao, Jin-Li Guo, Chao Fan, and Xue-Jiao Liu.
\newblock Individual and group dynamics in purchasing activity.
\newblock {\em arXiv.org}, 1010:--, October 2010.

\bibitem{topiAmos}
Alex Proekt, Jayanth~R Banavar, Amos Maritan, and Donald~W Pfaff.
\newblock Scale invariance in the dynamics of spontaneous behavior.
\newblock {\em Proceedings of the National Academy of Sciences},
  109(26):10564--10569, 2012.

\bibitem{Vazquez:2006zr}
A~Vazquez, J~G Oliveira, Z~Dezso, K~I Goh, I~Kondor, and A~L Barabasi.
\newblock {Modeling bursts and heavy tails in human dynamics}.
\newblock {\em Phys. Rev. E}, 73:036127, 2006.

\bibitem{edwards2007}
Andrew~M Edwards, Richard~A Phillips, Nicholas~W Watkins, Mervyn~P Freeman,
  Eugene~J Murphy, Vsevolod Afanasyev, Sergey~V Buldyrev, Marcos~GE da~Luz,
  Ernesto~P Raposo, H~Eugene Stanley, et~al.
\newblock Revisiting l{\'e}vy flight search patterns of wandering albatrosses,
  bumblebees and deer.
\newblock {\em Nature}, 449(7165):1044--1048, 2007.

\bibitem{Rybski2009}
D~Rynski, S~V Buldyrev, S~Havlin, F~Liljeros, and H~A Makse.
\newblock Scaling laws of human interaction activity.
\newblock {\em Proc. Natl. Acad. Sci. U.S.A.}, 106(31):12640--12645, 2009.

\bibitem{antenodo2009}
C~Anteneodo and DR~Chialvo.
\newblock Unraveling the fluctuations of animal motor activity.
\newblock {\em Chaos}, 19(3), 2009.

\bibitem{premananda2011}
P.~Indic, P.~Salvatore, C.~Maggini, S.~Ghidini, G.~Ferraro, R.~J. Baldessarini,
  and G.~Murray.
\newblock Scaling behavior of human locomotor activity amplitude: Association
  with bipolar disorder.
\newblock {\em PLoS ONE}, 6(5), 2011.

\bibitem{Barabasi:2005yq}
AL~Barabasi.
\newblock {The origin of bursts and heavy tails in human dynamics}.
\newblock {\em Nature}, 435(7039):207--211, 2005.

\bibitem{Kossinets:2006vn}
G~Kossinets and D.~J. Watts.
\newblock {Empirical Analysis of an Evolving Social Network}.
\newblock {\em Science}, 311(5757):--, January 2006.

\bibitem{Malmgren:2008ys}
R~Dean Malmgren, Daniel~B Stouffer, Adilson~E Motter, and Luis A~N Amaral.
\newblock {A Poissonian explanation for heavy tails in e-mail communication}.
\newblock {\em Proc. Natl. Acad. Sci. U.S.A.}, 105(47):18153--18158, 2008.

\bibitem{Eckmann:2004vn}
JP~Eckmann, E~Moses, and D~Sergi.
\newblock {Entropy of dialogues creates coherent structures in e-mail traffic}.
\newblock {\em Proc. Natl. Acad. Sci. U.S.A.}, 101(40):14333--14337, 2004.

\bibitem{Malmgren:2009rt}
R~Dean Malmgren, Daniel~B Stouffer, Andriana S. L.~O. Campanharo, and Luis
  A.~Nunes Amaral.
\newblock {On Universality in Human Correspondence Activity}.
\newblock {\em Science}, 325(5948):1696--1700, 2009.

\bibitem{Oliveira:2005fk}
Joao~Gama Oliveira and Albert-Laszlo Barabasi.
\newblock Human dynamics: {D}arwin and {E}instein correspondence patterns.
\newblock {\em Nature}, 437(7063):1251--1251, 10 2005.

\bibitem{zhao2011}
Zhao Zhi-Dan, Xia Hu, Shang Ming-Sheng, and Zhou Tao.
\newblock Empirical analysis on the human dynamics of a large-scale short
  message communication system.
\newblock {\em Chinese Physics Letters}, 28(6):068901, 2011.

\bibitem{quwang2011}
Shaocheng Qu, Qinqin Wang, and Li~Wang.
\newblock The statistical research of human dynamics on correspondence.
\newblock In {\em Modelling, Identification and Control (ICMIC), Proceedings of
  2011 International Conference on}, pages 282--284. IEEE, 2011.

\bibitem{stouffer2005comments}
DB~Stouffer, RD~Malmgren, and LAN Amaral.
\newblock Comments on ``the origin of bursts and heavy tails in human
  dynamics''.
\newblock {\em arXiv preprint physics/0510216}, 2005.

\bibitem{WuZhou2010}
Y~Wu, C~Zhou, J~Xiao, J~Kurths, and Schellnhuber~H J.
\newblock Evidence for a bimodal distribution in human communication.
\newblock {\em Proc. Natl. Acad. Sci. U.S.A.}, 107(44):18803--18808, 2010.

\bibitem{JohansenCommento2006}
A~Johansen.
\newblock Comment on {A.-L.} {B}arabasi, {N}ature 435 207-211 (2005).
\newblock {\em arXiv:physics/0602029v1}, 2006.

\bibitem{malmgremCritica2006}
D~B Stouffer, R~D Malmgren, and L~A~N Amaral.
\newblock Log-normal statistics in e-mail communication patterns.
\newblock {\em arXiv:physics/0605027v1}, 2006.

\bibitem{cobham}
A~Cobham.
\newblock Priority assignment in waiting line problems.
\newblock {\em J. Oper. Res. Soc. Amer.}, 2(1):70--76, 1954.

\bibitem{Abate1996}
J~Abate and W~Whitt.
\newblock Asymptotics for m/g/1 low-priority waiting-time tail probabilities.
\newblock {\em Queueing Systems}, 25:173--233, 1996.

\bibitem{Grinstein2008}
G~Grinstein and R~Linsker.
\newblock Power-law and exponential tails in a stochastic priority-based model
  queue.
\newblock {\em Phys. Rev. E}, 77:012101, 2008.

\bibitem{IAT1}
N.~Masuda, J.~S. Kim, and B.~Kahng.
\newblock Priority queues with bursty arrivals of incoming tasks.
\newblock {\em Phys. Rev. E}, 79:036106, Mar 2009.

\bibitem{IAT2}
Joris Walraevens, Thomas Demoor, Tom Maertens, and Herwig Bruneel.
\newblock Stochastic queueing-theory approach to human dynamics.
\newblock {\em Phys. Rev. E}, 85:021139, 2012.

\bibitem{blanchard2007}
Ph. Blanchard and M.-O. Hongler.
\newblock Modeling human activity in the spirit of {B}arabasi's queueing
  systems.
\newblock {\em Phys. Rev. E}, 75:026102, Feb 2007.

\bibitem{min2009}
Byungjoon Min, K-I Goh, and I-M Kim.
\newblock Waiting time dynamics of priority-queue networks.
\newblock {\em Physical Review E}, 79(5):056110, 2009.

\bibitem{cho2010}
Won-kuk Cho, Byungjoon Min, K-I Goh, and I-M Kim.
\newblock Generalized priority-queue network dynamics: Impact of team and
  hierarchy.
\newblock {\em Physical Review E}, 81(6):066109, 2010.

\bibitem{kim2010}
Kilhwan Kim and Kyung~C Chae.
\newblock Discrete-time queues with discretionary priorities.
\newblock {\em European Journal of Operational Research}, 200(2):473--485,
  2010.

\bibitem{crane2010}
Riley Crane, Frank Schweitzer, and Didier Sornette.
\newblock Power law signature of media exposure in human response waiting time
  distributions.
\newblock {\em Physical Review E}, 81(5):056101, 2010.

\bibitem{mailart2011}
Thomas Maillart, Didier Sornette, Stefan Frei, Thomas Duebendorfer, and
  Alexander Saichev.
\newblock Quantification of deviations from rationality with heavy tails in
  human dynamics.
\newblock {\em Physical Review E}, 83(5):056101, 2011.

\bibitem{saichev2010}
A~Saichev and D~Sornette.
\newblock Effects of diversity and procrastination in priority queuing theory:
  The different power law regimes.
\newblock {\em Physical Review E}, 81(1):016108, 2010.

\bibitem{jo2011}
Hang-Hyun Jo, Raj~Kumar Pan, and Kimmo Kaski.
\newblock Time-varying priority queuing models for human dynamics.
\newblock {\em Physical Review E}, 85(6):066101, 2012.

\bibitem{newman2009}
A~Clauset, C~R Shalizi, and M~E~J Newman.
\newblock Power-law distributions in empirical data.
\newblock {\em SIAM Review}, 51(4):661--703, 2009.

\bibitem{anteneodo2010}
C~Anteneodo, R~Dean Malmgren, and DR~Chialvo.
\newblock Poissonian bursts in e-mail correspondence.
\newblock {\em The European Physical Journal B}, 75(3):389--394, 2010.

\bibitem{Gabrielli2009}
A~Gabrielli and G~Caldarelli.
\newblock Invasion percolation and the time scaling behavior of a queuing model
  of human dynamics.
\newblock {\em Journal of Stat. Mech.-Theory and experiment}, FEB 2009.

\bibitem{Gabrielli2009-1}
A~Gabrielli and G~Caldarelli.
\newblock Invasion percolation on a tree and queueing models.
\newblock {\em Phys. Rev. E}, 79, Apr 2009.

\bibitem{Rousseau:2000tk}
B.~Rousseau and R.~Rousseau.
\newblock {LOTKA: A program to fit a power law distribution to observed
  frequency data}.
\newblock {\em Cybermetrics}, 4(1):4, 2000.

\bibitem{Conover:1972vs}
W~J Conover.
\newblock {A Kolmogorov Goodness-of-Fit Test for Discontinuous Distributions}.
\newblock {\em Journal of the American Statistical Association},
  67(339):591--596, September 1972.

\bibitem{Arnold:2011ty}
Taylor~B Arnold and John~W Emerson.
\newblock {Nonparametric Goodness-of-Fit Tests for Discrete Null
  Distributions}.
\newblock {\em The R Journal}, 3(2):34--39, 2011.

\bibitem{Kentsis:2006uq}
Alex Kentsis.
\newblock Correspondence patterns: Mechanisms and models of human dynamics.
\newblock {\em Nature}, 441(7092):E5--E5, 05 2006.

\bibitem{Vazquez:2005fk}
A~Vazquez.
\newblock Exact results for the {B}arabasi model of human dynamics.
\newblock {\em Phys. Rev. Letters}, 95:248701, 2005.

\bibitem{impact2007}
Alexei Vazquez.
\newblock Impact of memory on human dynamics.
\newblock {\em Physica A: Statistical Mechanics and its Applications},
  373:747--752, 2007.

\bibitem{impact2009}
JG~Oliveira and A~Vazquez.
\newblock Impact of interactions on human dynamics.
\newblock {\em Physica A: Statistical Mechanics and its Applications},
  388(2):187--192, 2009.

\bibitem{habit2010}
Yu~Jiao, YanHeng Liu, Jian Wang, and Jing Wang.
\newblock Model for human dynamics based on habit.
\newblock {\em Chinese Science Bulletin}, 55(24):2744--2749, 2010.

\bibitem{stasapower}
Sta{\v{s}}a Milojevi{\'c}.
\newblock Power law distributions in information science: making the case for
  logarithmic binning.
\newblock {\em Journal of the American Society for Information Science and
  Technology}, 61(12):2417--2425, 2010.

\bibitem{combinatorics}
P~Flajolet and R~Sedgewick.
\newblock {\em Analityc Combinatorics}.
\newblock Cambridge University Press (New York), 2008.

\end{thebibliography}

\end{document}